\documentclass[12pt]{article}

\pdfoutput=1
\usepackage{color}
\usepackage{epsfig, palatino}
\usepackage{pstricks,pst-node,pst-tree}
\usepackage{epic}
\usepackage{mathrsfs}
\usepackage{ae} 
\usepackage[T1]{fontenc}
\usepackage[ansinew]{inputenc}
\usepackage{amsmath}
\usepackage{amssymb}
\usepackage{graphicx}
\usepackage{ulem}
\usepackage{color}
\definecolor{darkblue}{cmyk}{0.9,0.9,0,0}
\usepackage[colorlinks=true,linkcolor=darkblue,citecolor=darkblue,urlcolor=darkblue]{hyperref}
\usepackage{cite}
\usepackage{hyperref}
\usepackage{wasysym}
\usepackage{varioref}
\usepackage{makeidx}
\usepackage[english]{babel}
\usepackage{simplewick}
\usepackage{array}
\usepackage{multirow}

\usepackage[font={small}]{caption}

\newcommand{\comment}[1]{}

\newcommand{\beq}{\begin{equation}}
\newcommand{\eeq}{\end{equation}}
\newcommand{\beqq}{\begin{equation*}}
\newcommand{\eeqq}{\end{equation*}}
\newcommand\beqa{\begin{eqnarray}}
\newcommand\eeqa{\end{eqnarray}}
\newcommand\beqaa{\begin{eqnarray*}}
\newcommand\eeqaa{\end{eqnarray*}}
\newcommand\bea{\begin{array}}
\newcommand\eea{\end{array}}

\newcommand{\nn}{\nonumber}

\newcommand{\neqa}{\nonumber\end{eqnarray}} 
\newcommand{\la}[1]{\label{#1}}

\newcommand{\eq}[1]{(\ref{#1})}

\renewcommand{\d}{\partial}

\newcommand{\<}{{\langle}}
\renewcommand{\>}{{\rangle}}

\newcommand{\re}{\relax{\rm I\kern-.18em R}}

\renewcommand{\sp}{p\hspace{-.40em}/}

\definecolor{darkgreen}{rgb}{0.0, 0.45, 0.0}

\newcommand{\Blue}[1]{{\color{blue}#1\color{black}}}
\newcommand{\Red}[1]{{\color{red}#1\color{black}}}
\newcommand{\Green}[1]{{\color{darkgreen}#1\color{black}}}

\def\XXint#1#2#3{{\setbox0=\hbox{$#1{#2#3}{\int}$}
\vcenter{\hbox{$#2#3$}}\kern-.5\wd0}}

\def\su2{{SU(2)}}

\def\[{\left[}
\def\]{\right]}

\def\({\left(}
\def\){\right)}
\def\[{\left[}
\def\]{\right]}

\def\<{\langle}
\def\>{\rangle}

\def\i2{\frac{i}{2}}

\def\spi{\relax{\rm \pi\kern-0.5em /}}
\def\sA{\relax{\rm A\kern-0.5em /}}
\def\sp{\relax{\rm p\kern-0.5em /}}
\def\sd{\relax{\rm \d\kern-0.5em /}}
\def\sk{\relax{\rm k\kern-0.5em /}}
\def\sn{\relax{\rm n\kern-0.5em /}}
\def\sl{\relax{\rm l\kern-0.5em /}}
\def\sP{\relax{\rm P\kern-0.7em /}}
\def\sBethe{\relax{\rm \Bethe\kern-0.5em /}}

\def\cF{{\cal F}}

\def\cF{{\cal F}}

\def\cD{{\cal D}}
\def\cR{{\cal R}}
\def\cO{{\cal O}}

\def\cP{{\cal P}}

\def\cW{{\cal W}}

\def\2F1{\,_2{\rm F}_1}
\def\sumint{\sum\hspace{-1.4em}\int}

\makeindex

\begin{document}

\thispagestyle{empty}

\renewcommand{\thefootnote}{\fnsymbol{footnote}}
\setcounter{page}{1}
\setcounter{footnote}{0}
\setcounter{figure}{0}

\begin{center}
$$$$
{\Large\textbf{\mathversion{bold}
OPE for all Helicity Amplitudes II. \\
Form Factors {and} Data Analysis
}\par}

\vspace{1.0cm}

\textrm{Benjamin Basso$^\text{\tiny 1,7}$, Jo\~ao Caetano$^\text{\tiny 2,3,4,5,7}$, Luc\'ia C\'ordova$^\text{\tiny 2,3,7}$, Amit Sever$^\text{\tiny 6}$ and Pedro Vieira$^\text{\tiny 2,7}$}
\\ \vspace{1.2cm}
\footnotesize{\textit{
$^\text{\tiny 1}$Laboratoire de Physique Th\'eorique, \'Ecole Normale Sup\'erieure, Paris 75005, France\\
$^\text{\tiny 2}$Perimeter Institute for Theoretical Physics,
Waterloo, Ontario N2L 2Y5, Canada\\
$^\text{\tiny 3}$Department of Physics and Astronomy \& Guelph-Waterloo Physics Institute, University of Waterloo, Waterloo, Ontario N2L 3G1, Canada\\
$^\text{\tiny 4}$Mathematics Department, King's College London, The Strand, London WC2R 2LS, UK\\
$^\text{\tiny 5}$Centro de F$\acute{\imath}$sica do Porto, Departamento de F$\acute{\imath}$sica e Astronomia,
Faculdade de Ci$\hat{e}$ncias da Universidade do Porto, Rua do Campo Alegre 687, 4169-007 Porto, Portugal\\
$^\text{\tiny 6}$School of Physics and Astronomy, Tel Aviv University, Ramat Aviv 69978, Israel\\
$^{\tiny 7}$ICTP South American Institute for Fundamental Research, IFT-UNESP, S\~ao Paulo, SP Brazil 01440-070 \\
}  
\vspace{4mm}
}

\par\vspace{1.5cm}

\textbf{Abstract}\vspace{2mm}
\end{center}
We present the general flux tube integrand for MHV and non-MHV amplitudes, in planar $\mathcal{N}=4$ SYM theory, up to a group theoretical rational factor. {We find that the MHV and non-MHV cases only differ by simple form factors which we derive.}
This information allows us to run the operator product expansion program for all sorts of non-MHV amplitudes and to test the recently proposed map with the so called charged pentagons transitions. Perfect agreement is found, on a large sample of non-MHV amplitudes, with the perturbative data available in the literature.

\noindent

\setcounter{page}{1}
\renewcommand{\thefootnote}{\arabic{footnote}}
\setcounter{footnote}{0}

\setcounter{tocdepth}{2}

 \def\nref#1{{(\ref{#1})}}

\newpage

\tableofcontents

\parskip 5pt plus 1pt   \jot = 1.5ex

\newpage

\section{Introduction} 

The main goal of the pentagon operator product expansion (POPE) program \cite{short} is to provide an explicit representation of \textit{any} gluon scattering amplitude in planar $\mathcal{N}=4$ Super Yang-Mills (SYM) theory at \textit{finite} coupling. 
With our current understanding of the POPE approach, such an answer will come in the form of an infinite sum, akin to a sort of partition function, over all possible excitations of the chromodynamic flux tube of the planar theory. More precisely, a scattering amplitude involving $n$ gluons is dual to a polygonal Wilson loop with $n$ edges which we decompose into $n-3$ successive fluxes/squares. In this picture, the transition from one flux to the next is induced by a pentagon operator. Hence, the amplitude can be identified with a form factor representation of a correlation function of the $n-4$ pentagon operators. The matrix elements of these operators are computed following a somehow standard integrable bootstrap \cite{short}. We shall refer to the summand arising in this form factor representation, as the \textit{POPE integrand}. This integrand depends, amongst other things, on the $n-5$ flux tube states which propagate in each of the $n-5$ middle squares
\beq\la{integrand} 
\centering\def\svgwidth{9cm}
\begingroup%
  \makeatletter%
  \providecommand\color[2][]{%
    \errmessage{(Inkscape) Color is used for the text in Inkscape, but the package 'color.sty' is not loaded}%
    \renewcommand\color[2][]{}%
  }%
  \providecommand\transparent[1]{%
    \errmessage{(Inkscape) Transparency is used (non-zero) for the text in Inkscape, but the package 'transparent.sty' is not loaded}%
    \renewcommand\transparent[1]{}%
  }%
  \providecommand\rotatebox[2]{#2}%
  \ifx\svgwidth\undefined%
    \setlength{\unitlength}{322.4bp}%
    \ifx\svgscale\undefined%
      \relax%
    \else%
      \setlength{\unitlength}{\unitlength * \real{\svgscale}}%
    \fi%
  \else%
    \setlength{\unitlength}{\svgwidth}%
  \fi%
  \global\let\svgwidth\undefined%
  \global\let\svgscale\undefined%
  \makeatother%
  \begin{picture}(1,0.53816405)%
    \put(0,0){\includegraphics[width=\unitlength]{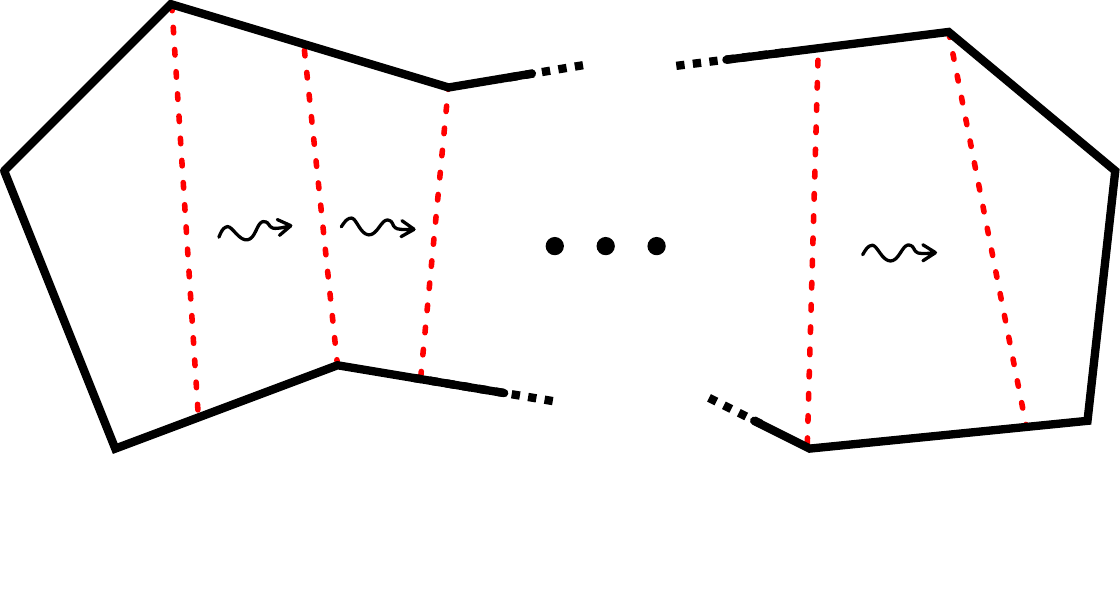}}%
    \put(0.18398705,0.35868412){\color[rgb]{0,0,0}\rotatebox{5.2715901}{\makebox(0,0)[lb]{\smash{${\bf\Psi}^{(1)}$}}}}%
    \put(0.73813974,0.34566026){\color[rgb]{0,0,0}\makebox(0,0)[lb]{\smash{${\bf\Psi}^{(n-5)}$}}}%
    \put(0.30011693,0.36811135){\color[rgb]{0,0,0}\makebox(0,0)[lb]{\smash{${\bf\Psi}^{(2)}$}}}%
    \put(0.04002023,0.0035211){\color[rgb]{0,0,0}\makebox(0,0)[lb]{\smash{$\texttt{integrand}= \texttt{integrand}({\bf\Psi}^{(1)},\dots,{\bf\Psi}^{(n-5)})$}}}%
  \end{picture}%
\endgroup%

\eeq
Such states are generically multi-particle states parametrized by a rapidity for each of the excitations. For instance, we could have, in the $i$-th square, a state with two gluons of positive helicity, a bound state of two gluons of negative helicity, one scalar and a pair of fermions:\footnote{Here $A,B,C$ and $D$ are $SU(4)$ R-charge indices and the indices on $F$ indicate the helicity of the gluonic excitation.}

\beq\la{state}
{\bf\Psi}^{(i)} =\left\{ F_1(u_1),F_1(u_2),F_{-2}(u_3) , \phi^{AB} (u_4) , \psi^C(u_5),\psi^D(u_6)\right\} \,.
\eeq
A finite coupling solution for scattering amplitudes in this gauge theory hinges on finding explicit expressions for the OPE integrand (\ref{integrand}) for any possible multi-particle flux tube state such as (\ref{state}). 

As described in detail below, what renders this seemingly gargantuan task feasible is a fortunate factorization of the OPE integrand into considerably simpler building blocks which we can analyse separately. 
It is perhaps worth mentioning from the get-go that this factorization is by no means obvious. It stands as another wonderful (but mysterious) $\mathcal{N}=4$ SYM gift. Were it not for it, one would hardly imagine bootstrapping the multi-particle contributions with ease. Indeed, except for the first few particles contributions, almost no explicit form factor summands for correlation functions in integrable theories are explicitly worked out. A notable exception is the $2d$ Ising model. 
The unexpected simplicity we are encountering in $\mathcal{N}=4$ SYM theory motivates its portrait as the \textit{Ising model of Gauge Theories}. 

The three building blocks into which the OPE integrand factorizes are dubbed the \textit{dynamical part}, the \textit{matrix part} and the \textit{form factor part},  
\beq\la{POPEintegrand}
\texttt{integrand} = (\texttt{dynamical part}) \times (\texttt{matrix part}) \times (\texttt{form factors part})\, ,
\eeq
with the latter form factors being non-trivial for non-MHV processes only. Here is what we know about these building blocks :

\begin{description}
\item[Dynamical part -] This is the part of the OPE integrand that is universally present, which applies to all cases, MHV or non-MHV, and which treats all flux tube excitations on a same footing, regardless of their quantum numbers / R-charges. It is the most dynamical component of the integrand, hence its name, and not surprisingly it exhibits the most complicated coupling dependence. Its overall form is however extremely simple since it is factorized into a product over elementary pentagon transitions linking the various flux tube excitations and a product over square measures and Boltzmann weights of each excitation. The geometry of the scattering amplitude, in particular, {only enters through these Boltzmann weights.} 
All the transitions, measures, energies and momentum appearing here are also rather universal. Most of them have already been spelled out, see e.g.~\cite{data,2pt,fusion,Belitsky:2014sla,Andrei1,Andrei2}, and all of them will be summarized in this paper. 
\item[Matrix part -] The matrix part takes care of the $SU(4)$ group theoretical factor of the integrand. It can only show up when flux tube excitations with R-indices are present, and is otherwise totally absent. (It is also trivial whenever there is only one way to distribute the R-indices.)
This component of the integrand has the distinguished feature of being a \textit{coupling independent} rational function of the particles' rapidities, with no obvious factorization. 
Taming this group theoretical factor is an interesting algebraic problem on its own~\cite{FrankToAppear} but is beyond the scope of this paper.
\item[Form factors part -] Lastly, we have the non-MHV form factors. They are only needed for non-MHV amplitudes, which are composed of so-called {\it charged} pentagon transitions~\cite{data,Andrei1,shortSuper}. Luckily, these form factors are not independent objects. Instead, we can construct them from their relation to the bosonic (or MHV) transitions with fermionic excitations frozen to zero momentum. Applying this logic, we will obtain their expressions for all excitations and transitions. The final result can then be tested against perturbative results as well as using self-consistency checks such as parity symmetry.
\end{description}
The main result of this paper is a complete recipe for writing the two coupling dependent factors in~(\ref{POPEintegrand}), that is the complete flux tube integrand up to the matrix part. This can be seen as $2/3$ of the full POPE program set above and is spelled out in section~\ref{OPEintegrand}. In section~\ref{data} we perform several perturbative checks of our POPE elements. The comparisons that we have performed test both the form factors and the dictionary proposed in~\cite{shortSuper} between N$^k$MHV amplitudes and charged pentagon sequences in a rather non-trivial way.

\section{The abelian part} \la{OPEintegrand}

In this section we present the expression for the abelian part of the POPE integrand. It captures by definition what remains of the full integrand after stripping out the matrix part. (In some cases, when there is just no matrix part, the abelian part is of course everything. This is the case for instance for states made out of gluons or their bound states, which are intrinsically abelian.) As explained in the introduction, {the abelian part is composed of the dynamical and form factors parts,}
\beq\la{abelian}
\texttt{abelian}= (\texttt{dynamical part}) \times   (\texttt{non-MHV form factors part}) \,.
\eeq
Conventionally, for MHV, only the dynamical part remains. For non-MHV, the latter remains the same, but form factors should be added to the story. These ones are not really independent and can be directly derived from suitable MHV processes, as we shall explain in this section.

\subsection{The dynamical part}

\begin{figure}[t]
\centering
\def\svgwidth{5cm}
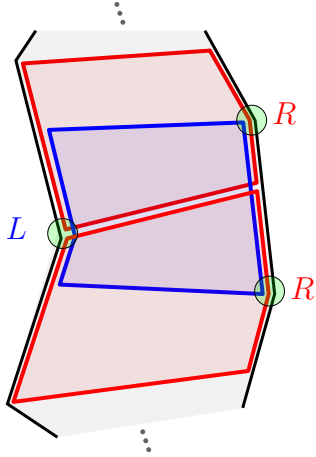
\caption{Every two successive pentagons in the POPE decomposition are flipped with respect to each other. Namely, if the cusp of one pentagon is pointing to the right then the next one is pointing to the left and so on, $\cdots\to \cP^\Red{R}\to \cP^\Blue{L}\to \cP^\Red{R}\to \cP^\Blue{L}\to\cdots $. } \label{Flipping}
\end{figure}

\begin{figure}[t]
\centering
\def\svgwidth{16cm}
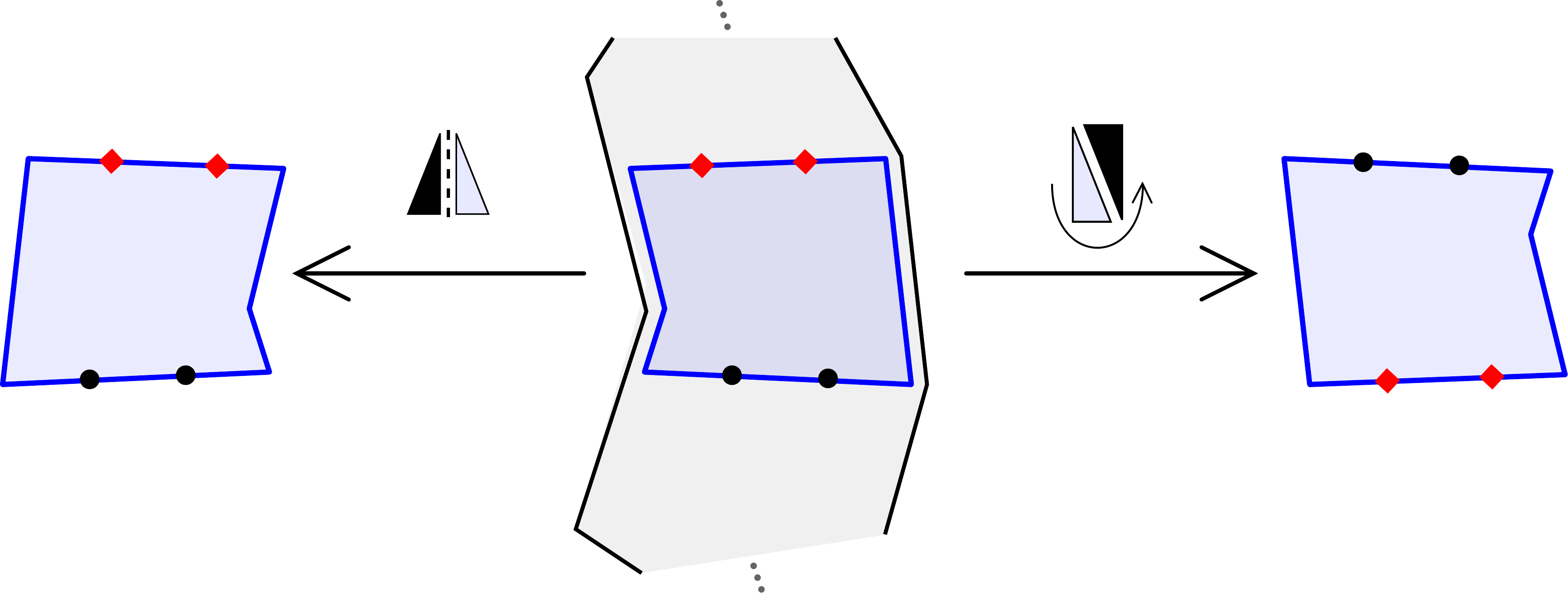
\caption{When inserting the resolution of the identity between each pentagon it is desirable to relate one sort of pentagon in figure \ref{Flipping} (say the ones with the cusp to the left) to the other kind to render things more uniform. To do so, in previous POPE works, a pentagon with a cusp to the left was related to a pentagon with its cusp to the right by a ``reflection", which maps the bottom to the bottom and the top to the top. In this paper, instead, we relate a pentagon with its cusp to the left to a pentagon with its cusp to the right by a ``rotation", which maps the bottom to the top and the top to the bottom. In principle both ought be equivalent. In practice, because of annoying minus signs inside some square roots in several pentagon transitions, the second convention is considerably more convenient as it avoids several ambiguities that would be present otherwise. In this new convention, after rotating the pentagon we relabel the associated rapidities as illustrated in this figure. This leads to an alternating sign $(-1)^j$ multiplying the flux tube space variables as written in (\ref{dynP}).} \label{FlippingRotating}
\end{figure}

We start with the main component. This one captures, in particular, the information about the geometry, i.e.~the cross ratios $\sigma_i, \tau_i, \phi_i$ of the polygon, and can be written as~\cite{short}
\beq\la{dynP}
\begin{aligned}
\texttt{dynamical part}&=P(\ \ 0\ \ |{\bf\Psi}^{(1)})\,\mu({{\bf\Psi}^{(1)}})\, e^{-E({{\bf\Psi}^{(1)}})\tau_1{+}ip({{\bf\Psi}^{(1)}})\sigma_1+im({{\bf\Psi}^{(1)}})\phi_1}\\
&\times P(\overline{\bf\Psi}^{(2)}|\overline{\bf\Psi}^{(1)})\,\mu({{\bf\Psi}^{(2)}})\, e^{-E({{\bf\Psi}^{(2)}})\tau_2{-}ip({{\bf\Psi}^{(2)}})\sigma_2+im({{\bf\Psi}^{(2)}})\phi_2}\\
&\times P({\bf\Psi}^{(2)}|{\bf\Psi}^{(3)})\,\mu({{\bf\Psi}^{(3)}})\, e^{-E({{\bf\Psi}^{(3)}})\tau_3{+}ip({{\bf\Psi}^{(3)}})\sigma_3+im({{\bf\Psi}^{(3)}})\phi_3}\\
&\times P(\overline{\bf\Psi}^{(4)}|\overline{\bf\Psi}^{(3)}) \dots\, ,
\end{aligned}
\eeq
where $E({\bf \Psi})$, $p({\bf \Psi})$ and $m({\bf \Psi})$ are the energy, momentum and angular momentum of the multi-particle state ${\bf \Psi}$. We have $n-5$ such states in total, in accordance with the number of middle squares in the tessellation, and for each of them we have a corresponding square measure~$\mu({{\bf\Psi}})$, see~\cite{short}. 
Finally, two consecutive squares with multi-particle states {$\bf\Phi$ and $\bf\Psi$} are connected by means of a pentagon transition $P({\bf\Phi}|{\bf\Psi})$ or $P(\overline{\bf\Psi}|\overline{\bf\Phi})$, where the bar stands for the state where all excitations are replaced by their conjugate and their order reversed. {The fact that each other pentagon appears with such reversed states is a direct consequence of the alternating nature of the pentagon tessellation as illustrated in figures \ref{Flipping} and \ref{FlippingRotating}. {The alternating signs multiplying the momenta of the states in consecutive middle squares have the same origin.}}


The factorization observed above is not a surprise and follows from symmetry considerations of the OPE. In contrast, the simplicity of $\mathcal{N}=4$ SYM theory starts to manifest itself as soon as we start exploring the multi-particle nature of the various {pieces}. What happens here is that all the above mentioned blocks factorize further into one- and two-particle blocks! To describe this factorization we introduce the notation $\Psi_n$ with $n=1,\dots,N$ to indicate the $n$-th excitation of the multi-particle state ${\bf \Psi}$. Then, the energy, momentum, angular momentum and measure all factorize into their single particle counterparts as\footnote{In \cite{short} the measure part also included combinatorial factors for identical excitations. These factors can instead be associated to the summation over the flux excitation that should be done in a way that avoids double counting.}
\beq
\mu({{\bf\Psi}})\,e^{-E({{\bf\Psi}})\tau\pm ip({{\bf\Psi}})\sigma+im({{\bf\Psi}})\phi}=  \prod_{n=1}^N \mu({{\Psi_n}})\,\exp\[{-E({{\Psi_n}})\tau\pm ip({{\Psi_n}})\sigma+im({{\Psi_n}})\phi}\]\, ,
\eeq
{where the sign $\pm= (-1)^{j+1}$ multiplying the momenta for states in the $j$-th middle square is a simple outcome of the conventions mentioned above, see figure \ref{FlippingRotating}.}
It is convenient to use a hatted measure $\hat \mu$ to denote collectively the measure and the accompanying {Boltzmann} factor, since these ones always come together. With this notation, the factorization we just described would simply read {$\hat\mu({{\bf\Psi}})=  \prod_{n} \hat\mu({{\Psi_n}})$}. 
Most importantly, we observe a similarly neat factorization for the pentagon transitions into fundamental 2-particle transitions~\cite{short,data,fusion,Belitsky:2014rba,Andrei2}\footnote{Let us stress again that formula~(\ref{Pfactorization}) only captures the dynamical part of the transition. In the case where ${\bf\Phi} = 0$ and $\bf\Psi =\phi\phi$, for example, we get from~(\ref{Pfactorization}) that $P_{\phi\phi}(0|u, v) = 1/P_{\phi|\phi}(u|v)$ while in \cite{2pt} we had $P_{\phi\phi}(0|u, v) = 1/(g^2(u-v+2i)(u-v+i)) \times 1/P_{\phi|\phi}(u|v)$ which differs by a rational prefactor and by the factor~$1/g^2$. The former rational factor is interpreted here as being part of the matrix part and thus discarded, while the power of $1/g^2$ is just absent because of our new normalization of $P_{\phi|\phi}$ (see appendix~\ref{alltransitions}).}
\beq\la{Pfactorization}
P({\bf\Phi}|{\bf\Psi})=\frac{\prod\limits_{n,m}P(\Phi_n|\Psi_m)}{\prod\limits_{n>n'}^{}P(\Phi_n|\Phi_{n'})\prod\limits_{m<m'}^{}P(\Psi_m|\Psi_{m'})} \,.
\eeq
The measures themselves are not independent from the pentagon transitions. On the contrary, they can be extracted from the decoupling pole present in 2-particle transitions involving identical in- and out- going particles,
\beq
\underset{v=u}{\text{Res}}\,P_{\Psi|\Psi}(u|v)=\frac{i}{\mu_{\Psi}(u)}\la{resPmu}\,.
\eeq

{We see that to fully describe the dynamical part all we need are the two-particle pentagon transitions between any pair of single particle excitations.
Most of them were already written down in the literature, see e.g.~\cite{data,2pt,fusion,Belitsky:2014sla,Andrei1,Andrei2}. It was found that, for any pair of excitations $\{\Psi,\Phi\}$, the pentagon transition takes the rather universal form}
\beq
P_{\Psi|\Phi}(u|v)^2=\mathcal{F}_{\Psi\Phi}(u,v) \frac{S_{\Psi\Phi}(u,v)}{S_{\star \Psi\Phi}(u,v)}\,, \label{struct}
\eeq
where $S_{\Psi\Phi}(u,v)$ is the scattering phase (in the symmetric channel) for the excitations $\Psi$ and $\Phi$ and $S_{\star \Psi\Phi}(u,v)$ is its mirror counterpart.\footnote{When $\Psi$ is a gluon or scalar excitation, the mirror S-matrix is given by analytically continuing the physical one in the standard way, $S_{\star \Psi\Phi}(u,v)=S_{\bar{\Psi}\Phi}(u^{\gamma},v)$. For fermions, the lack of a mirror transformation $u^\gamma$ renders the relation between the mirror S-matrix and the physical one less straightforward. It involves a so-called anomalous mirror transformation as detailed in \cite{2pt}, see e.g. appendix A.4 therein.} 
The functions $\mathcal{F}_{\Psi\Phi}(u,v)$ are simple functions of the rapidities, involving eventually the 
Zhukowski variables. In table \ref{squaredptab} we present our ansatz for these functions for all pairs of excitations. 

\begin{table}[t]
\beq
\begin{aligned} 
&\mathcal{F}_{\phi  F}(u|v)  =1\, ,\\
&\mathcal{F}_{\phi  \psi}(u|v) = -\frac{1}{(u-v+\frac{i}{2})} \, , \\
&\mathcal{F}_{\phi  \phi}(u|v) = \frac{1}{(u-v)(u-v+i)} \, , \\
&\mathcal{F}_{FF}(u|v) = \frac{(x^{+}y^{+}-g^2)(x^{+}y^{-}-g^2)(x^{-}y^{+}-g^2)(x^{-}y^{-}-g^2)}{g^2 x^{+}x^{-}y^{+}y^{-}(u-v)(u-v+i)}\, , \\
&\mathcal{F}_{F\psi}(u|v) = -\frac{(x^{+}y-g^2)(x^{-}y-g^2)}{g\sqrt{x^{+}x^{-}}y (u-v+\frac{i}{2})}\, , \\
&\mathcal{F}_{F  \bar{\psi}}(u|v) = -\frac{g\sqrt{x^{+}x^{-}}y(u-v+\frac{i}{2})}{(x^{+}y-g^2)(x^{-}y-g^2)} \, , \\
&\mathcal{F}_{F\bar{F}}(u|v) = \frac{g^2x^{+}x^{-}y^{+}y^{-}(u-v)(u-v+i)}{(x^{+}y^{+}-g^2)(x^{+}y^{-}-g^2)(x^{-}y^{+}-g^2)(x^{-}y^{-}-g^2)}\, , \\
&\mathcal{F}_{\psi\psi}(u|v) = -\frac{(xy-g^2)}{\sqrt{gxy}(u-v)(u-v+i)}\, , \\
&\mathcal{F}_{\psi\bar{\psi}}(u|v) = -\frac{\sqrt{gxy}}{(xy-g^2)} \, ,
\end{aligned}
\eeq
\caption{{Summary of prefactors for all twist-one squared transitions, with $x = x(u), y = x(v)$ and $x = \frac{1}{2}(u+\sqrt{u^2-4g^2})$ the Zhukowski variable. {They agree with those found in the literature up to minor redefinitions (see appendix A for details).}}}  \la{squaredptab}
\end{table}

In the end, to evaluate the transition we still need to take the square-root of (\ref{struct}) which poses some ambiguity on the branch choice. {This ambiguity can be partially fixed through comparison with data or with help of some reasonable normalization conditions as done in appendix~\ref{alltransitions}. Explicit expressions for all the transitions and measures are also given in this appendix (altogether with the transitions involving bound states of gluons and small fermions \cite{2pt}).}

{In concluding, we recall that there are two main dynamical inputs behind these ans\"atze. The most important (and still mysterious) one is the \textit{fundamental relation}
\beq\la{fundamentalrelation}
P_{X|Y}(u|v) = \pm S_{XY}(u,v)P_{Y|X}(v|u)\, ,
\eeq
which comes with a minus sign whenever both $X$ and $Y$ are fermionic. Combined with the factorization property~(\ref{Pfactorization}) it guarantees that general transitions fulfill proper Watson and decoupling equations. The other essential constraint is dubbed the \textit{mirror axiom} which states that 
\beq\la{mirror}
P_{X|Y}(u^{-\gamma}|v) = P_{ \bar Y|X}(v|u)\, ,\quad \text{or more generally}\quad P_{X|{\bf\Psi}}(u^{-\gamma}|{\bf v}) = P_{{\bf\bar\Psi}|X}({\bf v}|u)\,,
\eeq
where $-\gamma$ denotes the inverse mirror rotation and with $\bf v$ a set of spectator rapidities.%
\footnote{Technically, $X$ in this equation is restricted to be a gluonic or a scalar excitation. This is because, as explained in \cite{2pt}, the mirror rotation for the fermions is of a more exotic type, mapping the fermions into higher-twist excitations.}
Combined with~(\ref{fundamentalrelation}) the mirror axiom can be used to argue for the ansatz~(\ref{struct}) as well as to solve for the prefactor $\mathcal{F}$. Finally, it is worth stressing that the solution to such bootstrap axioms is by no means unique and comparison with perturbative data is therefore crucial in backing up our proposals. This shall be discussed at length below.}

\subsection{Charged transitions and form factors}\la{sec2point2}

In the OPE framework, one can cover all helicity amplitudes at once by introducing a super pentagon transition~\cite{shortSuper}
\beq
\mathbb{P} = \mathcal{P} + \chi^{A}\mathcal{P}_{A} +  \chi^{A}\chi^{B}\mathcal{P}_{AB} + \chi^{A}\chi^{B}\chi^{C}\mathcal{P}_{ABC} + \chi^{A}\chi^{B}\chi^{C}\chi^{D}\mathcal{P}_{ABCD}\, ,
\eeq
where $\chi^A$ is a Grassmann parameter in the fundamental representation of the $SU(4)$ R-symmetry group, and where $\mathcal{P}_{A_1\ldots A_k}$, or $P^{[k]}$ for short, is the so-called charged transition transforming in the $k$-th antisymmetric product. In the same way that arbitrary MHV amplitudes can be described by sequences of bosonic pentagons $\mathcal{P}$, one can generate all non-MHV amplitudes by gluing super pentagon transitions together, e.g. 
\beq
\cW =\underbrace{\cP\circ\cP\circ\dots\circ\cP}_{\rm MHV}+\chi_1^1\chi_1^2\chi_1^3\chi_1^4\,\underbrace{\cP_{1234}\circ\cP\circ\dots\circ\cP}_{\rm NMHV}+\chi_1^1\chi_1^2\chi_1^3\chi_2^4\,\underbrace{\cP_{123}\circ\cP_4\circ\dots\circ\cP+}_{\rm NMHV}\dots\, \la{superloopinPs}
\eeq
That this is enough information for recovering the many components of the super Wilson loop is not a priori obvious, since the $\chi$ parameters here are attached to the pentagons in the sequence, and not to all possible edges of the loop. However, the missing `degrees of freedom' are somewhat superfluous since controlled by supersymmetry, and, as explained in~\cite{shortSuper}, the $\chi$ components are as many supersymmetry independent components as necessary to fix a general super amplitude.

The physics that takes place on the flux tube is essentially the same regardless of whether some of the pentagons are charged or not. What can possibly differ is the R-charge flow throughout the pentagon evolution. Since the R-charge dependence was factored out into the matrix part at the very beginning, one can ask if the abelian part proposed in~(\ref{Pfactorization}) can also be applied as it stands to these charged processes. The answer turns out to be positive up to a minor modification: the inclusion of the so-called non-MHV form factors, as sketched in~(\ref{abelian}). The need for these form factors is not a novelty and was previously stressed in \cite{data, Andrei2} from the study of certain components of the NMHV hexagon. To pave the way to our general discussion, let us start by reviewing briefly, on a simple example, why these form factors are needed at all, or equivalently why is the dynamical part not enough for describing the abelian part of non-MHV amplitudes.

Consider the $\chi$-component $\cP_{1234}\circ\cP$, or equivalently $\cP\circ\cP_{1234}$, of an NMHV hexagon. From R-charge conservation the excitations allowed on these transitions are the same as in the bosonic MHV case $\cP\circ\cP$. As such, at twist zero we have the vacuum, at twist one the positive and negative helicity gluons, $F$ and $\bar{F}$, etc. Despite this similarity, one does not expect the transitions, integrands, and full amplitude to be the same for the two processes, since e.g. the MHV process treats symmetrically positive and negative helicity gluons while the non-MHV one does not. (This is also immediately confirmed at weak coupling by looking at the corresponding amplitudes.)  At the level of the POPE integrand for a single gluon,
\beqa
\cP\circ\cP\ \ \ \  &=&1+\int\frac{du}{2\pi}\,\hat\mu_{F}(u)\qquad\,+\int\frac{du}{2\pi}\,\hat\mu_{\bar{F}}(u)\qquad+\ldots\, ,\nn\\
\cP_{1234}\circ\cP &=&1+\int\frac{du}{2\pi}\,\hat\mu_{F}(u)\,f(u)+\int\limits\frac{du}{2\pi}\,\hat\mu_{\bar{F}}(u)\,\bar f(u)+\ldots\, ,\la{gluonicexample}\\
\cP\circ\cP_{1234}&=&1+\int\frac{du}{2\pi}\,\hat\mu_{F}(u)\,\bar f(u)+\int\frac{du}{2\pi}\,\hat\mu_{\bar{F}}(u)\,f(u)+\ldots\, ,\nn
\eeqa
this difference follows from the fact that the gluons are either produced or annihilated in the presence of a charged transition $P^{[4]}$ in the NMHV cases and it results in the gluonic form factors $f$ and $\bar f$. In other words, the dynamical part described before must be completed with the knowledge of these form factors, which in the present cases simply read~\cite{data}
\beq\label{simplefact}
f(u) = \frac{x^{+}x^{-}}{g^2} = 1/\bar{f}(u)\, ,
\eeq
where $x^{\pm} = x(u\pm \frac{i}{2})$ with $x(u) = (u+\sqrt{u^2-4g^2})/2$ the Zhukowski map of the rapidity $u$. Note that such factor cannot be absorbed in the matrix part both because the matrix part is, by definition, independent of the coupling and because in this abelian case there is no matrix part.

\begin{figure}[t]
\centering
\def\svgwidth{12cm}
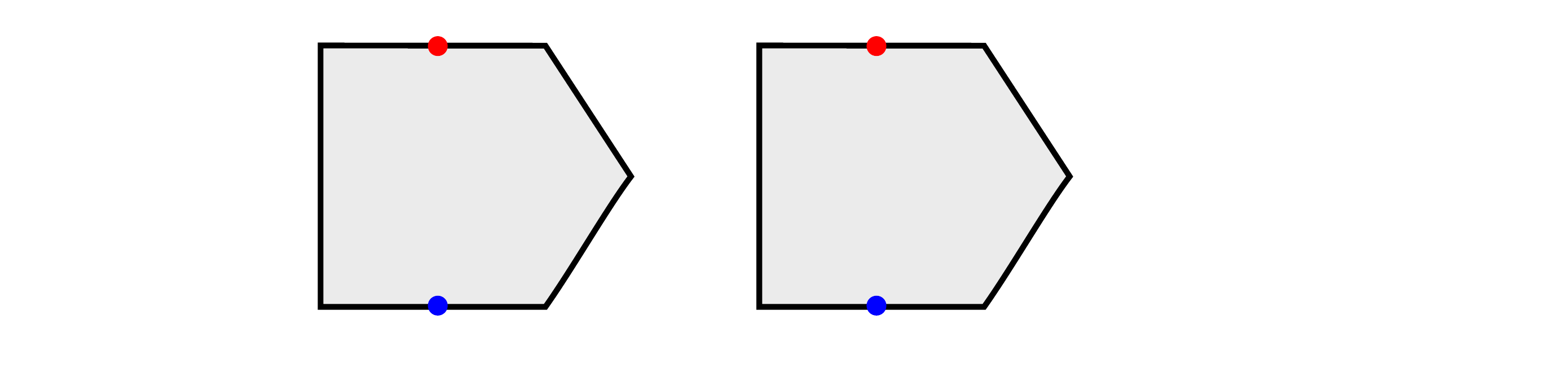
\caption{The net effect of charging a pentagon is to dress each excitation on its edges by a corresponding form factor. This one is attached to the field that creates or annihilates the corresponding excitation in the bottom to top evolution picture. The picture above illustrates our conventions, with a state $\Psi$ at the bottom being created by a field $\Psi$ and a state $\Phi$ at the top being annihilated by the field $\bar{\Phi}$. The form factors associated to these fields are then $\(h_{{\Psi}}\)^r$ and $\(h_{\bar{\Phi}}\)^r$, with $r$ the R-charge of the charged pentagon, or equivalently, the number of times we act with a supercharge $\cal Q$.} \label{eq21}
\end{figure}

Our main proposal is that the same structure persists for generic transitions. Namely, the effect of charging a pentagon, ignoring the matrix part, is to dress the abelian part by elementary form factors associated to each excitation present on the pentagons. More specifically, we propose that the charged version of~(\ref{Pfactorization}) reads
\beq\la{factorizedff}
P^{[r]}({\bf \Phi}|{\bf\Psi})= g^{\frac{r(r-4)}{8}}\times\[\prod_{i} \(h_{\Phi_i}\)^r\times \prod_{i} \(h_{\bar{\Psi}_i}\)^r\]\times P({\bf \Phi}|{\bf\Psi})\, ,
\eeq
where ${\bf{\Phi}}=\{\Phi_i\}$ is the incoming set of excitations at the bottom square of the pentagon, ${\bf\Psi}=\{\Psi_i\}$ the outgoing set of excitations at the top, and $r =0, 1, \ldots, 4$ the amount of R-charge carried by the pentagon. The rules of the game are extremely simple. The result is factorized and for each excitation, or more precisely for each field creating the excitation at the bottom or conjugate field annihilating it at the top, we associate a form factor. The form factor can thus be thought of as being attached to the field and represents the net effect of charging the transition, as illustrated in figure~\ref{eq21}.

We immediately verify that the general rule~(\ref{factorizedff}) properly reduces to~(\ref{gluonicexample}) in the case of a single gluon. Indeed, as mentioned earlier, the form factor $f$ above accounts for the difference between a gluon $F$ produced on top of a charged and an uncharged pentagon, i.e.
\beq
f(u) = P^{[4]}_{0|F}(0|u)/P_{0|F}(0|u)\, .
\eeq
By convention, or equivalently by applying~(\ref{Pfactorization}) blindly, $P(0|u) = 1$, while~(\ref{factorizedff}) gives us
\beq
P^{[4]}_{0|F}(0|u) = g^{0}\times (h_{\bar{F}}(u))^4\times 1\, ,
\eeq 
that is
\beq\label{GluonFF1}
f(u) = (h_{\bar{F}}(u))^4 \qquad \Leftrightarrow \qquad h_{\bar{F}}(u) = \left(\frac{x^{+}x^{-}}{g^2}\right)^{\frac{1}{4}}\, .
\eeq
Similarly, we would read that
\beq\la{GluonFF2}
h_{F}(u) = \left(\frac{g^2}{x^{+}x^{-}}\right)^{\frac{1}{4}}\, .
\eeq

The remaining questions are what are the form factors for the other excitations, why is the form factor for a composite state a product of elementary ones, and why is~(\ref{factorizedff}) valid at all? The quick answers are that, due to their expected simple dependence (see~(\ref{simplefact})), form factors are easily extracted from data analysis, as done in~\cite{data}, and that their factorized form is consistent with all the constraints the charged transitions must fulfill, as shall be discussed in section~\ref{consistency} below. However, one can do much better than that and actually \textit{derive} the rule~(\ref{factorizedff}) \textit{directly} from the uncharged transitions~(\ref{Pfactorization}). The important observation~\cite{shortSuper} is that charging a pentagon is the same as acting with a supersymmetry generator on one of its edges, as we will now explain.

To start with, we recall that one can view the pentagon transitions as form factors for a pentagon operator acting on the flux tube Hilbert space of states
\beq
\centering\def\svgwidth{7cm}
\begingroup%
  \makeatletter%
  \providecommand\color[2][]{%
    \errmessage{(Inkscape) Color is used for the text in Inkscape, but the package 'color.sty' is not loaded}%
    \renewcommand\color[2][]{}%
  }%
  \providecommand\transparent[1]{%
    \errmessage{(Inkscape) Transparency is used (non-zero) for the text in Inkscape, but the package 'transparent.sty' is not loaded}%
    \renewcommand\transparent[1]{}%
  }%
  \providecommand\rotatebox[2]{#2}%
  \ifx\svgwidth\undefined%
    \setlength{\unitlength}{395.06643066bp}%
    \ifx\svgscale\undefined%
      \relax%
    \else%
      \setlength{\unitlength}{\unitlength * \real{\svgscale}}%
    \fi%
  \else%
    \setlength{\unitlength}{\svgwidth}%
  \fi%
  \global\let\svgwidth\undefined%
  \global\let\svgscale\undefined%
  \makeatother%
  \begin{picture}(1,0.358117)%
    \put(0,0){\includegraphics[width=\unitlength]{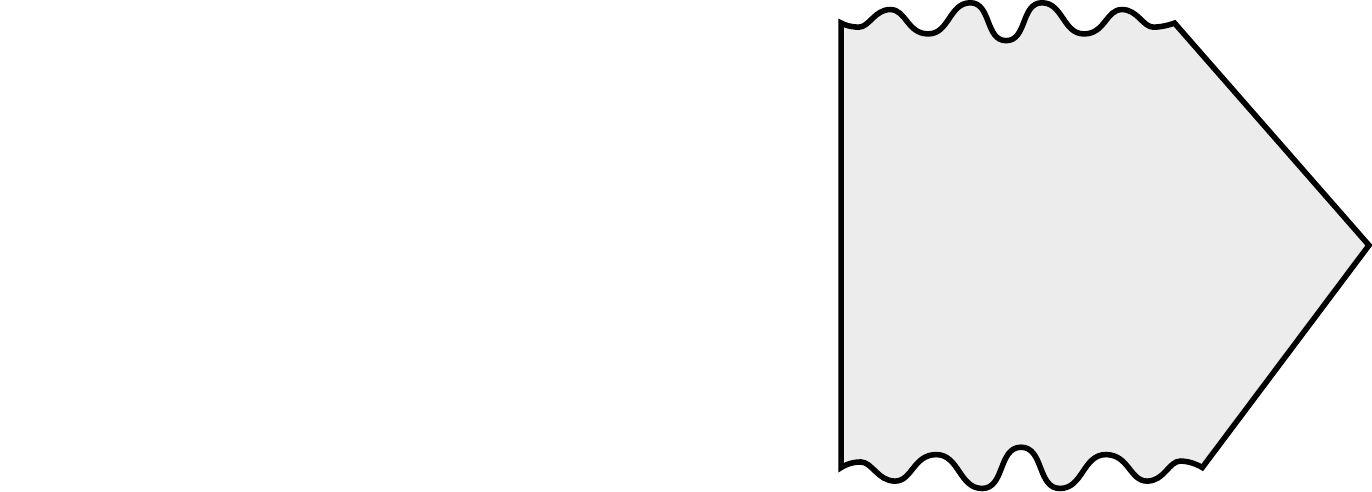}}%
    \put(0.72865325,0.04160603){\color[rgb]{0,0,0}\makebox(0,0)[lb]{\smash{${\bf \Psi}$}}}%
    \put(0.7246033,0.26840333){\color[rgb]{0,0,0}\makebox(0,0)[lb]{\smash{$\overline{\bf\Phi}$}}}%
    \put(-0.00093954,0.17316934){\color[rgb]{0,0,0}\makebox(0,0)[lb]{\smash{$P({\bf\Psi}|{\bf\Phi})= \<{\bf\Phi}|\cP|{\bf\Psi}\>=$}}}%
  \end{picture}%
\endgroup%

\eeq
As explained in \cite{shortSuper}, all we need to do in order to add a unit of R-charge to a pentagon $\cP$ is to act with a supersymmetry generator ${\cal Q}$ on the bottom state or, equivalently, with $-{\cal Q}$ on the top state. Here $\cal Q$ is the unique supercharge that commutes with $\cP$ and is represented by $\d_\chi$ on the super loop (\ref{superloopinPs}), see~\cite{shortSuper}. This leads to the relation :
\beq\la{PQ}
\centering\def\svgwidth{10cm}
\begingroup%
  \makeatletter%
  \providecommand\color[2][]{%
    \errmessage{(Inkscape) Color is used for the text in Inkscape, but the package 'color.sty' is not loaded}%
    \renewcommand\color[2][]{}%
  }%
  \providecommand\transparent[1]{%
    \errmessage{(Inkscape) Transparency is used (non-zero) for the text in Inkscape, but the package 'transparent.sty' is not loaded}%
    \renewcommand\transparent[1]{}%
  }%
  \providecommand\rotatebox[2]{#2}%
  \ifx\svgwidth\undefined%
    \setlength{\unitlength}{571.06645508bp}%
    \ifx\svgscale\undefined%
      \relax%
    \else%
      \setlength{\unitlength}{\unitlength * \real{\svgscale}}%
    \fi%
  \else%
    \setlength{\unitlength}{\svgwidth}%
  \fi%
  \global\let\svgwidth\undefined%
  \global\let\svgscale\undefined%
  \makeatother%
  \begin{picture}(1,0.24774701)%
    \put(0,0){\includegraphics[width=\unitlength]{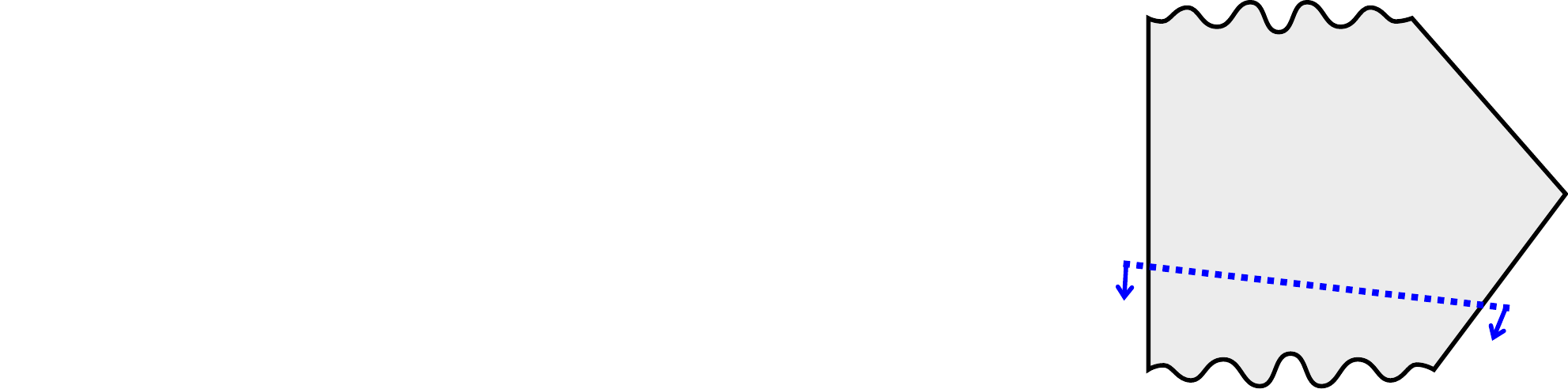}}%
    \put(0.81228102,0.02878325){\color[rgb]{0,0,0}\makebox(0,0)[lb]{\smash{${\bf \Psi}$}}}%
    \put(0.80947925,0.18568267){\color[rgb]{0,0,0}\makebox(0,0)[lb]{\smash{$\overline{\bf\Phi}$}}}%
    \put(-0.00064997,0.11979935){\color[rgb]{0,0,0}\makebox(0,0)[lb]{\smash{$P_\Blue{A}({\bf\Psi}|{\bf\Phi})= \<{\bf\Phi}|\cP_{\Blue{A}}|{\bf\Psi}\>=\<{\bf\Phi}|\cP\Blue{{\cal Q}_A}|{\bf\Psi}\>=$}}}%
    \put(0.82387179,0.093619){\color[rgb]{0,0,0}\rotatebox{-7.33139248}{\makebox(0,0)[lb]{\smash{$\Blue{{\cal Q}_A}$}}}}%
  \end{picture}%
\endgroup%

\eeq
Importantly, supersymmetry generators are realized on the flux as zero momentum fermions \cite{AldayMaldacena}. That is, to act on the state $|{\bf \Psi}\>$ with a supercharge ${\cal Q}_A$, we add to it a fermion and then send the momentum of that fermion to zero. This directly links the charged transitions to their un-charged counterparts, allowing us to extract all information about the former from the latter.

Let us show more precisely how this works. Consider for example the transition from the vacuum at the bottom to a single fermion $\bar{\psi}(u)$ at the top. Such transition is only possible if we equip the transition with R-charge as in the charged transition $P_A(0|u)=\<\bar{\psi}(u)|\cP_A|0\>$. To obtain the latter transition from an uncharged one we can start with a similar fermion $\bar{\psi}(v)$ at the bottom, i.e.~from $P_{\bar{\psi}\bar{\psi}}(v|u)$. We now wish to take the limit in which the momentum of the fermion $\bar{\psi}(v)$ goes to zero, so that it becomes a supersymmetry generator acting at the bottom. In appendix \ref{Qcommutator} we show carefully how the zero momentum limit should be taken and in particular what is the proportionality factor. We find that 
\beq\la{Qtopsi}
{\cal Q}|0\>=\sqrt{\Gamma_\text{cusp}\over 2g}\lim_{p\to0}|p\>=\lim_{v\to\infty}\sqrt{{\Gamma_\text{cusp}\over 2ig}\,\frac{d\check{v}}{dp_{\bar\psi}}\mu_{\bar\psi}(\check v)}\,|\bar{\psi}(\check v)\>\,,
\eeq
{where, following the notations of~\cite{2pt}, the `check mark' on top of the rapidity $v$, i.e.~$\check{v}$, indicates that the analytical continuation to the rapidity plane neighbouring the zero momentum point (reached at $v = \infty$) has been done (see~\cite{2pt} for further details).} Using this prescription, as well as the large $v$ behaviours given in appendix~\ref{zmf}, we conclude that
\beq\la{toformfactor}
 P_{0|\bar{\psi}}^{[1]}(0|u)=\lim_{v\to\infty}\sqrt{{\Gamma_\text{cusp}\over 2ig}\,\frac{d\check v}{dp_{\bar\psi}}\mu_{\bar\psi}(\check v)}\times P_{\bar\psi|\bar\psi}(\check v|u)
 =g^{-\frac{3}{8}}\({g\over x(u)}\)^{1\over4}\,,
\eeq
where the upper label in $P^{[1]}$ indicates the amount of R-charge or equivalently the number of $\chi$'s carried by the pentagon. The fermionic creation transition (\ref{toformfactor}) is obviously of the type~(\ref{factorizedff})%
\footnote{Recall that following~(\ref{Pfactorization}) we are working in a convention where the un-charged creation transition of any excitation $X$ is trivial. With the $\chi$-labelling this reads $P_{0|X}^{[0]}(0|u)=1$.}
\beq
P_{0|\bar{\psi}}^{[1]}(0|u)\equiv g^{-\frac{3}{8}}\times h_\psi(u)\times P_{0|\bar \psi}(0|u) =g^{-\frac{3}{8}}\times h_\psi(u)\, ,
\eeq
with 
\beq 
h_{\psi}(u) =({g/x})^{1/4} \la{hPSI}
\eeq the form factor for a single antifermion. As expected it shows a simple dependence on the rapidity $u$ of the excitation, once expressed in terms of the Zhukowski variable $x = x(u)$, as found earlier for the gluons.

It is not so much difficult to include more excitations in the top and check the factorization of the form factors. Suppose, for instance, that we start with a multi-particle transition involving fermions and send the momentum of one of them to zero as prescribed by (\ref{Qtopsi}). As a result of the multi-particle factorization (\ref{Pfactorization}), the multi-particle form factors must factorize as well.%
\footnote{To see this in full generality one also needs the matrix part~\cite{FrankToAppear}.}

We can see this at work on simple examples, using the same $P^{[1]}$ procedure as before. {For instance, we can at no cost consider the same fermion creation transition with a gluonic excitation $F_{a}(w)$ added on the top. This yields}
\beq\la{hforgluon}
P_{0|\bar\psi F_{a}}^{[1]}(0|u,w)
=\lim_{v\to\infty}\sqrt{{\Gamma_\text{cusp}\over 2ig}\,\frac{d\check v}{dp_{\bar\psi}}\mu_{\bar\psi}(\check v)} \times P_{\bar\psi|\bar\psi F_{a}}(\check v|u, w)\,,
\eeq
where, again, the upper index indicates that this new transition is taking place on top of a pentagon carrying one unit of R-charge. It exactly differs from the chargeless transition defined through~(\ref{Pfactorization}) by the form factors. Indeed, using the factorization of the dynamical part (\ref{Pfactorization}) together with~(\ref{toformfactor}), (\ref{hPSI}) and~(\ref{zmf}), we verify that
\beq
P_{0|\bar\psi F_{a}}^{[1]}(0|u, w) = g^{-\frac{3}{8}}h_{\psi}(u)h_{F_{-a}}(w) P_{0|\bar\psi F_{a}}(0|u, w) \, ,
\eeq
with
\beq\la{bsff}
h_{F_{a}}(u)=\left[\frac{g^2}{x\left(u+\frac{i a}{2}\right) x\left(u-\frac{i a}{2}\right)}\]^{(\text{sign}\, a)/4}\, ,
\eeq
in agreement with~(\ref{GluonFF1}) and~(\ref{GluonFF2}) for $a = -1$ and $a=+1$, respectively.

{Instead of adding matter to the fermion $\bar{\psi}$ at the top, one can imagine replacing it by a pair $\phi\psi$ or a triplet $\psi\psi\psi$ {(always with the small fermion $\bar \psi(\check v)$ at the bottom)}
and hence access to the as-yet-unknown $\phi$ and {$\bar\psi$} form factors. A proper analysis would require introducing a matrix part, whose main role is to project the pair/triplet to the $SU(4)$ channel with one unit of R-charge. However, in both cases, the matrix part plays no role as far as the form factors are concerned.%
\footnote{For completeness, we have indeed that the matrix part is $\propto 1/(u-w+3i/2)$ and $\propto 1/\prod_{i<j}(u_{i}-u_{j}+i)$ for the two cases at hand, i.e.~for a state $\phi(u)\psi(w)$ and $\psi(u_{1})\psi(u_{2})\psi(u_{3})$ at the top, respectively. Because it shows no dependence at all on the rapidity $v$ of the fermion $\bar{\psi}(\check{v})$ at the bottom, it cannot contribute to the form factors.}
One can thus proceed without knowing its explicit form and directly relate the $\phi$ and {$\bar\psi$} form factors to the large $v$ behaviours of the $P_{\bar{\psi}|\phi}(\check{v}|u)$ and $P_{\bar{\psi}|\psi}(\check{v}|u)$ transitions. Using expressions in appendix~\ref{zmf} one gets the system of equations
\beq
\begin{aligned}
h_{\phi}(u)h_{\bar{\psi}}(w) &= g^{3/8}\lim_{v\to\infty}\sqrt{{\Gamma_\text{cusp}\over2ig}\,\frac{d\check v}{dp_{{\bar\psi}}}\mu_{{\bar\psi}}(\check v)}\times P_{\bar\psi|\phi}(\check v|u)P_{\bar\psi|\psi}(\check v|w) = \left(\frac{x(w)}{g}\right)^{1/4}\, , \\
\prod\limits_{i=1}^{3}h_{\bar{\psi}}(u_{i}) &= g^{3/8}\lim_{v\to\infty}\sqrt{{\Gamma_\text{cusp}\over2ig}\,\frac{d\check v}{dp_{{\bar\psi}}}\mu_{{\bar\psi}}(\check v)}\times \prod\limits_{i=1}^{3}P_{\bar\psi|\psi}(\check v|u_{i}) = \left(\frac{x(u_{1})x(u_{2})x(u_{3})}{g^3}\right)^{1/4}\, , \\
\end{aligned}
\eeq
whose only reasonable solution is}
\beq\la{sff}
h_\phi(u)=1\,, \qquad h_{\bar{\psi}}(u) = (x/g)^{1/4}\, .
\eeq

{Equations~(\ref{hPSI}),~(\ref{bsff}), and~(\ref{sff}) finalize our proposal for the charged transitions~(\ref{factorizedff}). A couple of consistency checks for it will be given in the next sub-section. One easy test can actually be run immediately.} It comes from the physical requirement that a pair of conjugate excitations should decouple on a charged transition exactly as they do in the un-charged case. That is, the \textit{square limit} \eqref{resPmu} must remain the same on a charged transition.\footnote{Put differently, the propagation on the square is diagonal, so it cannot be charged.} Including the form factors, this condition translates into
\beq
\underset{v=u}{\text{Res}}\,P_{\Phi|\Phi}(u|v)\,h_{\Phi}\,(u)h_{\bar\Phi}(v)=\frac{i}{\mu_{\Phi}(u)}\,,
\eeq
which enforces
\beq
h_{\Phi}(u)h_{\bar\Phi}(u)=1\,. \label{squarelim}
\eeq
This relation is easily seen to be satisfied.

\begin{figure}[t]
\centering
\def\svgwidth{11cm}
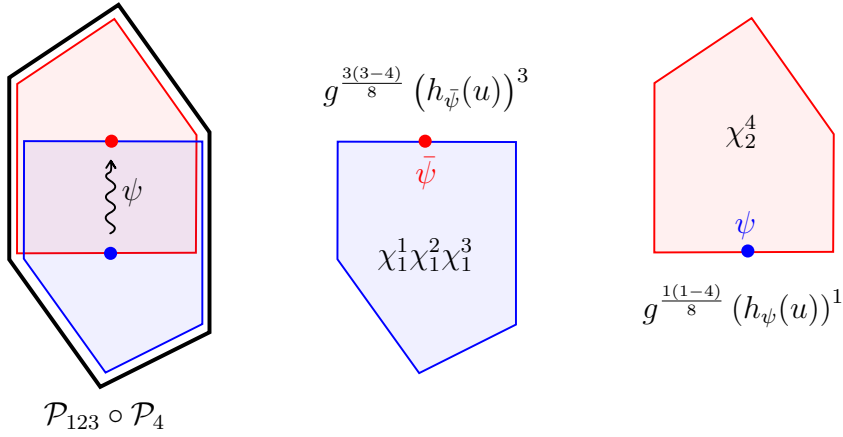
\caption{Leading twist transition for the hexagon component $\cP_{123}\circ\cP_4$. In this example, we assign the form factor $\(h_{\bar\psi}(u)\)^3$ to the bottom (blue) pentagon and $\(h_\psi(u)\)^1$ to the top (red) pentagon.} \label{ffconv}
\end{figure}

In the end our form factors are all simply given in terms of Zhukowski variables. Putting them together, for a polygon with $n$ edges, gives us the full form factors part in~(\ref{POPEintegrand}) as
\beq \label{ffactors}
\texttt{form factors part} =\prod\limits_{i=0}^{n-5}g^{r_i(r_i-4)\over8} \(h_{{\bf \Psi}^{(i)}}\)^{r_i}\(h_{{\bf \bar\Psi}^{(i+1)}}\)^{r_{i}}\, ,
\eeq
{where the index $i$ on top of the matter fields refers to the $i$'th square, with $i=0$ being the first one at the very bottom and $i=n-4$ the last one at the very top,}%
\footnote{We recall that the states in the very bottom ($i=0$) and very top $(i=n-4)$ squares are both vacuum with measure and form factors equal to one.} while the \textit{same} index $i$ in $r_i$ refers to the {$i$'th pentagon transition between states ${\bf \Psi}^{(i)}$ and ${\bf \Psi}^{(i+1)}$}, with $r_{i}$ units of R-charge. {Note in particular that each excitation in a given square is assigned two form factors -- one for each of the pentagons that overlap on this particular square.} {A simple example is depicted in figure \ref{ffconv}.}

{As a summary, we can now write the OPE decomposition for a polygon with $n$ edges in a rather compact form. Up to the matrix part of course, we have
\beq
\boxed{
\mathbb{P}\circ\mathbb{P}\circ\ldots\circ\mathbb{P}\rvert_X=\sumint\,\prod\limits_{i=0}^{n-5} \hat\mu_{{\bf \Psi}^{(i)}}\,   g^{r_i(r_i-4)\over8}\(h_{{\bf \Psi}^{(i)}}\)^{r_{i}-r_{i-1}} \, P^{R/L}({{\bf \Psi}^{(i)}|{\bf \Psi}^{(i+1)}})\, , } \la{FULLintegrand}
\eeq
with $X$ a choice of $\chi$ component, with $r_{i}$ Grassmann variables $\chi$'s in pentagon~$i$, and with $P^{R/L}({{\bf \Psi}^{(i)}|{\bf \Psi}^{(i+1)}}) = P({{\bf \Psi}^{(i)}|{\bf \Psi}^{(i+1)}})$ or $P({\overline{{\bf \Psi}}^{(i+1)}|\overline{{\bf \Psi}}^{(i)}})$ for $i$ even or odd. }

In what follows we shall perform two sort of checks of our proposal. The first kind of checks -- with which we will conclude this section -- are internal self-consistency checks of the POPE proposal. The second sort of checks concern explicit comparison against perturbative data and are the main focus of section \ref{data}.  

\subsection{Consistency checks}\label{consistency}

In this section we shall present two consistency checks of the non-MHV form factors presented above. Namely, we will first see that they are consistent with parity and finally observe that they are compatible with {the general axioms for multiparticle transitions.}

\subsubsection*{Parity}

We have proposed in \cite{shortSuper} a simple realization of parity within the POPE approach. In a given pentagon with $r$ units of $R-$charge, the action of parity conjugates its $R-$charge, i.e.  $r\rightarrow4-r$, which means flipping the chirality of the external particles that are being scattered, but also the chirality of the flux tube excitations,  namely $\psi \leftrightarrow \bar{\psi}$ and  $F \leftrightarrow \bar{F}$. This can be achieved by simply flipping the signs of the angles $\phi$ appearing in the square propagation factor of the OPE integrand.

In more concrete terms, parity symmetry establishes the following relation between the OPE components
\beq
\mathbb{P}\circ\mathbb{P}\circ\ldots\circ\mathbb{P}\rvert_X=\(\mathbb{P}\circ\mathbb{P}\circ\ldots\circ\mathbb{P}\rvert_{\bar X}\)\rvert_{{\phi}\rightarrow-{\phi}}\la{parityOPE2}\,.
\eeq
where $\bar{X}$ is the complement of the component $X$, namely
\beq
\bar{X} = \left.\int \prod_{i=1}^{n-4} d^4 \chi_i\, e^{\,\sum_{i=1}^{n-4}\bar{\chi_i} \chi_i} X\right|_{{\bar\chi}\rightarrow{{\chi}}}\,.
\eeq 
{For example, we can relate an NMHV component with $X=\chi_1^1\chi_1^2\chi_1^3\chi_2^4$ to an N$^{n-5}$MHV with its complement $\bar X=\chi_1^4\chi_2^1\chi_2^2\chi_2^3\dots\chi_{n-4}^1\chi_{n-4}^2\chi_{n-4}^3\chi_{n-4}^4$. For parity to be a symmetry of the super Wilson loop, the POPE decomposition should be invariant under $r\rightarrow4-r$ together with a flip of the chirality of all the flux tube excitations, namely
\beqa\la{parityequal}
&&\sumint\,\prod\limits_{i=0}^{n-5}\hat\mu_{{\bf \Psi}^{(i)}}\,   g^{r_i(r_i-4)\over8}\(h_{{\bf \Psi}^{(i)}}\)^{r_{i}-r_{i+1}}\,P^{R/L}({{\bf \Psi}^{(i)}|{\bf \Psi}^{(i+1)}})\,
\\
&=&\left.\sumint\,\prod\limits_{i=0}^{n-5}\hat\mu_{\bar{\bf \Psi}^{(i)}}\,   g^{(r_i-4)r_i\over8}\(h_{\bar{\bf \Psi}^{(i)}}\)^{(4-r_{i})-(4-r_{i+1})}\,P^{R/L}({\overline{\bf \Psi}^{(i)}|\overline{\bf \Psi}^{(i+1)}})\,\right|_{\phi_i\to-\phi_i}\nn\,,
\eeqa
up to an unphysical relative normalization.} Note that the overall power of $g$ is parity invariant by itself. Now, the pentagon transitions and measures change at most by a sign under this transformation (namely when fermions are involved) but they do not distinguish the chirality of the flux tube excitations otherwise. Finally, using the relation \eqref{squarelim} we see that the form factor part is also invariant and thus parity is nicely satisfied.

\subsubsection*{Bootstrap axioms}

{Given the set of elementary transitions (\ref{struct}), the multiparticle ansatz (\ref{Pfactorization}) solves minimally for the Watson, mirror and decoupling axioms that the multiparticle transitions are conjectured to satisfy \cite{short,data,2pt,fusion}. Any other solution to the same bootstrap equations would differ from~(\ref{Pfactorization}) by a so-called Castillejo-Dalitz-Dyson (CDD) factor. The charged transitions introduced in (\ref{factorizedff}) and their associated non-MHV form factors are particular examples of non-minimal solutions featuring totally factorized CDD factors. Indeed, as we shall now explain, the non-MHV form factors are such as to make the charged transitions (\ref{factorizedff}) a valid solution to the multiparticle bootstrap equations.}

{As already mentioned, the main feature of the prefactor in (\ref{factorizedff}) is that it is totally factorized. As such it is oblivious to the ordering of the excitations at both the bottom and the top of the pentagon. It then does not affect the Watson equation encoding the transformation property of the transitions upon permutation of excitations in the past or in the future.%
\footnote{Note however that the non-MHV prefactor is \textit{not} invariant under the exchange of a bottom and a top excitation. It cannot therefore be seen as a CDD factor for the elementary transitions (\ref{struct}) without spoiling their fundamental relations to the flux tube S-matrix (\ref{fundamentalrelation}). This corrects a misleading statement made in earlier versions of this paper. Examples of (factorized) CDD factors for the elementary transitions are given at the end of the appendix \ref{alltransitions}. These are redefinitions of the individual pentagon transitions that preserve the fundamental and mirror relations, eq.~(\ref{fundamentalrelation}) and first eq.~in~(\ref{mirror}), respectively. They also preserve the decoupling pole (\ref{resPmu}), when applicable, but renormalize the square measures.}
A factorized prefactor is also automatically regular and non-vanishing when bottom and top rapidities coincide. The decoupling pole and the decoupling condition for the multiparticle transitions are thus preserved, as already stressed around eq.~(\ref{squarelim}).} It remains to check the \textit{mirror axiom} (\ref{mirror}) which was introduced when bootstrapping the pentagon transitions involving gluons or scalars.  
For this mirror relation to be satisfied by the charged transitions, the form factor should obey the relation
\beq
h_{\Phi}(u^{-\gamma})=h_{\bar\Phi}(u)\,.
\eeq
This relation is immediately observed for scalars while for gluons it follows from the relation $x^{\pm}(u^{-\gamma})=g^2/x^{\pm}(u)$.\footnote{Beware that mirror transformation takes different form for scalars and gluons~\cite{MoreDispPaper,data}.}  In fact, the validity of both axioms directly follows from the relation between the charged transition and a zero momentum fermion (\ref{Qtopsi}) (taken to be one of the $\bf\Psi$ excitations in (\ref{mirror}) for example).

\section{Comparison with data}\la{data}

We would like now to test our proposal \eqref{FULLintegrand} for the {POPE} integrand against available data at weak coupling. Historically, this comparison was absolutely instrumental in unveiling the general ansatz for the form factors.  

The data we use is extracted from the package \cite{Bourjaily:2013mma}, which generates non-MHV amplitudes at tree level for any number of particles. The same package also yields one loop amplitudes but for the purpose of this paper we restrict our attention to tree-level checks only. 

In the {POPE} we have essentially five different types of elementary excitations, $F, \psi, \phi, \bar{\psi}, \bar{F}$, and fifteen different pairings of them into transitions. 
These two numbers are in correspondence with the five independent NMHV hexagons and fifteen independent NMHV heptagons, respectively. At leading twist, the hexagons essentially probe the measures and the form factors. The heptagons on the other hand probe all building blocks of the integrand and, in particular, the pentagon transitions involving mixed types of particles. In this section we will confront our predictions against both hexagons and heptagons thus probing all these ingredients at once.

At the same time, these checks also provide us with a strong test of the map between N$^k$MHV amplitudes and charged pentagons proposed in \cite{shortSuper}. This one maps any N$^k$MHV amplitude, as specified by choosing an $\eta$-component of the super loop, to a very precise linear combination of the OPE friendly $\chi$-components (\ref{superloopinPs}) and vice-versa. It is the latter $\chi$-components that admit a neat OPE interpretation and thus offer direct access to the various pentagon transitions.

\begin{figure}[t]
\centering
\def\svgwidth{16cm} 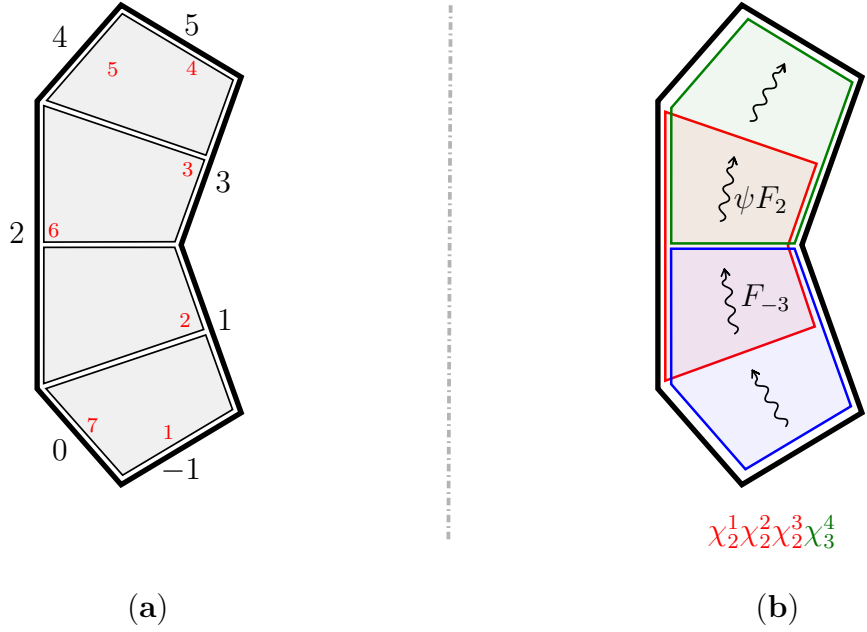
\caption{ ({\bf a}) {The POPE} friendly edge labelling used in this paper (big black outer numbers) versus the more conventional cyclic labelling (small red inner numbers) for the heptagon. ({\bf b}) The NMHV heptagon process analysed in this section. }
\label{FIGheptagon}
\end{figure}

To illustrate this point, let us consider a random $\eta$-component, say the NMHV heptagon component $\mathcal{W}^{(-1,-1,1,2)}$ multiplying the monomial $\eta_{-1}^A \eta_{-1}^B \eta_{1}^C \eta_{2}^D \epsilon_{ABCD}$ (see figure \ref{FIGheptagon}.{\bf a} for the convention of the edge labelling). According to our discussion, we expect it \textit{not} to have an obvious OPE expansion and indeed this is precisely what we find. To see it we extract {this} tree-level component\footnote{Actually, the package extracts the ratio function component $\cR^{(i,j,k,l)}=\cW^{(i,j,k,l)}/\cW_{\text{MHV}}$, but at tree level they are the same.} from the package \cite{Bourjaily:2013mma} as
\begin{center}
\verb|evaluate@superComponent[{1,2},{3},{},{},{},{4},{}]@treeAmp[7,1]| 
\end{center}
and, {to make it into a weight free quantity, multiply it by the weights $(({\bf -1})_1)^2\, ({\bf 1})_1\, ({\bf 2})_2$ with $(\bold{i})_j$ the weight of the  twistor $Z_i$ in the $j^{th}$ pentagon (see \cite{shortSuper} for more details),}
which we can express in terms of {OPE} variables using the twistors in~\cite{data}.%
\footnote{As is often the case, while this is the correct mathematical procedure, the naive evaluation of $\mathcal{W}^{(-1,-1,1,2)}$ with the twistors in \cite{data} would yield the very same result since, in this case, the weights simply evaluate to~$1$.} We denote this properly normalized heptagon component as $\mathbb{W}^{(-1,-1,1,2)}$. In terms of OPE variables we have
\beq
\mathbb{W}^{(-1,-1,1,2)}=\frac{e^{-{\sigma_1}-\frac{i {\phi_1}}{2}}}{1+e^{-2\tau_1}}=e^{-{\sigma_1}-\frac{i {\phi_1}}{2}}-e^{-{\sigma_1}-2 {\tau_1}-\frac{i {\phi_1}}{2}} + \cO(e^{-4 {\tau_1}})\,,\label{hepexp}
\eeq
which clearly defies any reasonable OPE interpretation!\footnote{To start with, it simply does not depend on the OPE variables $\tau_2,\sigma_2,\phi_2$ at all, as if only the vacuum were propagating in the second square of this heptagon. Even if we were to accept that, other puzzles would immediately appear when interpreting the large $\tau_1$ expansion: the first term looks like a twist zero contribution -- like the vacuum does -- but with a non-trivial $\sigma_1$ and $\phi_1$ dependence -- contrary to the vacuum. There is no natural candidate for what anything like this would be. Also suspicious is the fact that only even twists show up once we expand the result out at large $\tau_1$. 
Moreover terms in the near collinear expansion shout trouble: their dependence in $\sigma_1$ is so simple that they would not have any sensible Fourier transform into momentum space. In short: this component is as weird as it could be from an OPE perspective.}

Of course, this does \textit{not} mean that we cannot describe this component within the POPE approach. On the contrary, as explained in \cite{shortSuper}, once using the inverse map we can express any component as a linear component of the nice $\chi$-components which in turn we can describe at any loop order within the POPE. In this case, using the expression (21) of \cite{shortSuper} we would find
\beq
\begin{aligned}
 \mathbb{W}^{(-1,-1,1,2)} =\; &\cP_{12}\circ\cP\circ\cP_{34}\; e^{-\sigma _1-\sigma _2-\tau _2+\frac{i \phi _1}{2}}+\\
&\cP_{12}\circ\cP_3\circ\cP_{4}\; \left( e^{-2 \sigma _1-\tau _1-\tau _2-\frac{i \phi _1}{2}-\frac{i \phi _2}{2}}+e^{-\sigma _1-\tau _2+\frac{i \phi _1}{2}-\frac{i \phi _2}{2}}+e^{-\sigma _1-\sigma _2-2 \tau _2+\frac{i \phi _1}{2}+\frac{i \phi _2}{2}}-\right.\\
& \left. \hspace{69pt} e^{-\sigma _1-\sigma _2+\frac{i \phi _1}{2}+\frac{i \phi _2}{2}} \right)+\\
&\cP_{12}\circ\cP_{34}\circ\cP\; \left(-e^{-2 \sigma _1-\tau _1-\frac{i \phi _1}{2}}-e^{-\sigma _1-\sigma _2-\tau _2+\frac{i \phi _1}{2}+i \phi _2}-e^{-\sigma _1+\frac{i \phi _1}{2}}\right)+\\
&\cP_{123}\circ\cP\circ\cP_4\; \left( e^{-2 \sigma _1-\tau _2-\frac{i \phi _2}{2}}+e^{-\sigma _1-\sigma _2-\tau _1+i \phi _1+\frac{i \phi _2}{2}}+e^{-\sigma _2+\frac{i \phi _2}{2}} \right)+\\
&\cP_{123}\circ\cP_{4}\circ\cP\; \left(e^{-\sigma _1-\tau _1-i \phi _1}+e^{-\sigma _1-\tau _1+i \phi _1}+e^{-\sigma _2-\tau _2+i \phi _2}+e^{-\sigma _1-\sigma _2-\tau _1-\tau _2+i \phi _1+i \phi _2}+\right.\\
& \left. \hspace{69pt}e^{-2 \sigma _1-2 \tau _1}-e^{-2 \sigma _1}+1\right)+ \\
&\cP_{1234}\circ\cP\circ\cP \; \left( e^{-2 \sigma _1-\tau _1+\frac{i \phi _1}{2}}+e^{-\sigma _1-\frac{i \phi _1}{2}}\right)\,.\label{monster}
\end{aligned}
\eeq
{It is amusing to see how this precise linear combination of $\chi$-components, each with a nice OPE expansion, combines into the component (\ref{hepexp}) without any obvious OPE picture.}
Conversely, and perhaps less trivially, according to \cite{shortSuper}, a generic $\chi$-component is a precise linear combination of several $\eta$- components. {These are the components that we shall directly confront against the integrability inspired predictions in the following subsections.}

\subsection{NMHV Hexagon}\la{NMHVhex}
Due to R-charge conservation, different NMHV components will support different flux tube transitions. The hexagon -- made out of two pentagons -- is the simplest case where we can clearly see this at work. For the components $\cP_{1234}\circ\cP$ and $\cP\circ\cP_{1234}$, encountered before, the excitations flowing in the middle square should form an R-charge singlet (for example we could have the vacuum, any bound state of gluons, a pair $\psi_1\bar\psi_{234}$, etc.). For $\cP_{123}\circ\cP_{4}$ ($\cP_{1}\circ\cP_{234}$) we should have excitations with the same total R-charge as for a fermion (anti-fermion), that is the state should be in the fundamental (antifundamental) representation of $SU(4)$. Finally for $\cP_{12}\circ\cP_{34}$ we need the same R-charge as for a scalar, i.e. a state in the vector representation of the R-symmetry group. These five components form a basis over which we can expand any other component \cite{shortSuper}. 

\begin{figure}[t]
\centering
\def\svgwidth{16.5cm} 
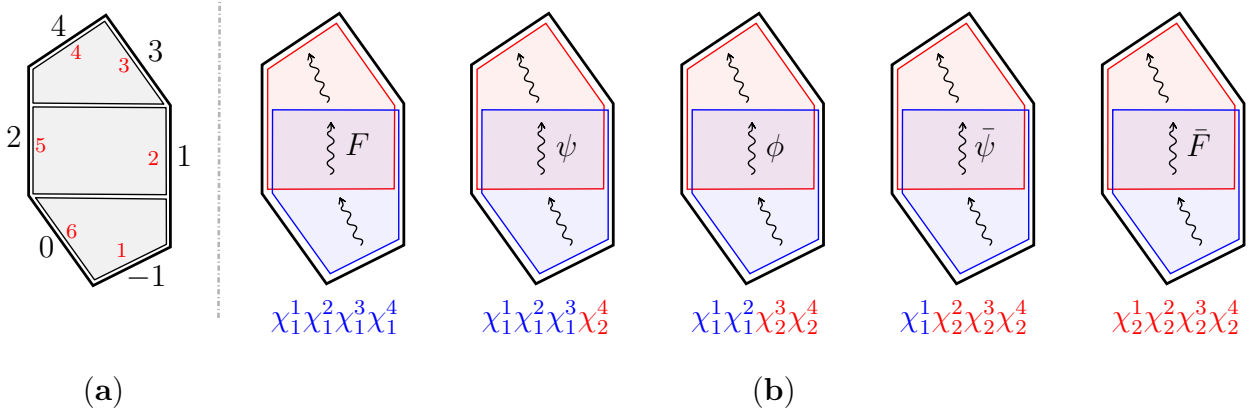
\caption{ ({\bf a}) OPE friendly edge labelling used in this paper (big black outer numbers) versus the more conventional cyclic labelling (small red inner numbers) for the hexagon. ({\bf b})~The five components of the NMHV hexagon and the corresponding excitations at twist one. In the first and last component we have written the particle appearing first in perturbation theory.}
\label{FIGhexagon}
\end{figure}

We can write explicitly the OPE integrand for these nice components. At leading twist, they are (see figures \ref{ffconv} and \ref{FIGhexagon})
\beqa\label{hexagons}
\cP_{1234}\circ\cP &=&1+\int\limits_{\mathbb{R}} \frac{du}{2\pi}\, \hat\mu_{F_1}(u)\,(h_{F_{-1}}(u))^4\,(h_{F_1}(u))^0+\int\limits_{\mathbb{R}} \frac{du}{2\pi}\, \hat\mu_{F_{-1}}(u)\,(h_{F_{1}}(u))^{4}\,(h_{F_{-1}}(u))^{0}+\dots\,, \nn \\
\cP_{123}\circ\cP_{4}&=&g^{-\frac{3}{4}}\int\limits_{\mathcal{C}} \frac{du}{2\pi} \, \hat\mu_{\psi}(u)\, (h_{\bar\psi}(u))^{3}\,h_{\psi}(u)+\dots\,  \label{comppsi}\,,\\
\cP_{12}\circ\cP_{34}&=&g^{-1}\int\limits_{\mathbb{R}} \frac{du}{2\pi} \,\hat\mu_{\phi}(u)\, (h_{\phi}(u))^2 (h_{\phi}(u))^2+\dots\,,\nn \\
\cP_{1}\circ\cP_{234}&=&g^{-\frac{3}{4}}\int\limits_{\mathcal{C}}\frac{du}{2\pi} \,\hat\mu_{\bar\psi}(u)\,h_{\psi}(u)\, (h_{\bar\psi}(u))^3+\dots\,, \nn\\
\cP\circ\cP_{1234}&=&1+\int\limits_{\mathbb{R}}  \frac{du}{2\pi}\,\hat\mu_{F_{-1}}(u)\,(h_{F_{1}}(u))^0\, (h_{F_{-1}}(u))^4+\int\limits_{\mathbb{R}} \frac{du}{2\pi}\,\hat\mu_{F_{1}}(u)\, (h_{F_{-1}}(u))^0\, (h_{F_{1}}(u))^4+\dots\,. \nn
\eeqa
Although in the first and last line we have the same allowed excitations, the form factors break the symmetry between the positive and negative helicity gluons. In particular, the first terms in $\cP_{1234}\circ\cP$ and $\cP\circ\cP_{1234}$ appear at tree level and the last terms are delayed to two loops, as confirmed from data. 

As thoroughly described in \cite{2pt}, the contour of the integration $\mathcal{C}$ for the fermions is over a two-sheeted Riemann surface and can be conveniently splitted into two contributions: the large and small sheet contours. The large sheet contour is performed over the real axis with a small positive imaginary part whereas the small sheet contour is a counter-clockwise half-moon on the lower complex plane. It is then natural to treat these contributions independently as coming from a large fermion $\psi_L$ and a small fermion $\psi_S$.
In appendix \ref{APPsmall} we provide the explicit formulae for the analytic continuation of the  pentagon transitions from the large to the small sheet.

Let us now study in detail the component $\cP_{123}~\circ~\cP_4$ at leading twist as an illustration of a check against data (see figure \ref{ffconv}). The OPE integral splits into two terms corresponding to the large and small fermion contributions
\beq
\cP_{123}\,\circ\,\cP_{4}=\, g^{-\frac{3}{4}} \int\limits_{\mathcal{C}_\text{large}} \frac{du}{2\pi}\, \hat\mu_{\psi_L}(u)\, (h_{\bar\psi_L}(u))^3\,h_{\psi_L}(u)+g^{-\frac{3}{4}}\int\limits_{\mathcal{C}_\text{small}} \frac{du}{2\pi} \, \hat\mu_{\psi_S}(u)\, (h_{\bar\psi_S}(u))^3\,h_{\psi_S}(u) \,.
\eeq
In this case the contour of integration $\mathcal{C}_\text{small}$ does not enclose any singularity, resulting in a vanishing contribution of the second term {at any value of the coupling \cite{2pt}}. We are only left with the first term that should be integrated slightly above the real axis. Using the explicit expressions for the measure and form factor at leading order in the coupling, we obtain
\beq
\cP_{123}\circ\cP_{4}= e^{-\tau+i\phi/2} \int\limits_{\mathbb{R}+i0}\frac{du}{2\pi}\, \frac{-i \pi}{\sinh (\pi u)} e^{2iu \sigma}+\cO(g^2)\,.
\eeq
According to the map worked out in \cite{shortSuper}, this component should relate to a component of the super amplitude as follows
\beq
\cP_{123}\circ\cP_{4} = \((\bold{-1})_1\)^3 (\bold{4})_2\, \mathcal{W}^{(-1,-1,-1,4)}\,,
\eeq
where, as before, the pre-factor $(\bold{i})_j$ stands for the weight of the twistor $Z_i$ in the $j^{th}$ pentagon. In order to extract this component we use the aforementioned package by running the following line in Mathematica
\begin{center}
\verb|evaluate@superComponent[{1,2,3},{},{},{4},{},{}]@treeAmp[6,1]|
\end{center}
still using the twistors given in \cite{data}.%
\footnote{Note that one should convert between the OPE friendly and the cyclic labelling of the edges. Figure~\ref{FIGhexagon}.\textbf{a} shows both labellings for the hexagon.}
Upon expanding the outcome at leading order in the twist and taking into account the weights, we obtain a perfect match validating our conjecture for this particular transition.

The same type of checks are straightforward to generalize to the rest of the components or to one loop level using the same package \cite{Bourjaily:2013mma}. These probe the expressions for the form factors presented in \eqref{ffactors} at weak coupling. We have also verified the correctness of our conjectures beyond leading twist when also the pentagon transitions start to play a role.  
In the next section, we probe them more directly using the NMHV heptagon.

\subsection{NMHV Heptagon}\la{NMHVhep}

The NMHV heptagon is the appropriate laboratory to test the pentagon transitions involving all possible pairings of the fundamental excitations. In particular, it is the first polygon where the transitions between excitations in different squares arise. All the POPE building blocks take now part in and consequently they are all scrutinised. 

\begin{figure}
\centering
\def\svgwidth{16cm}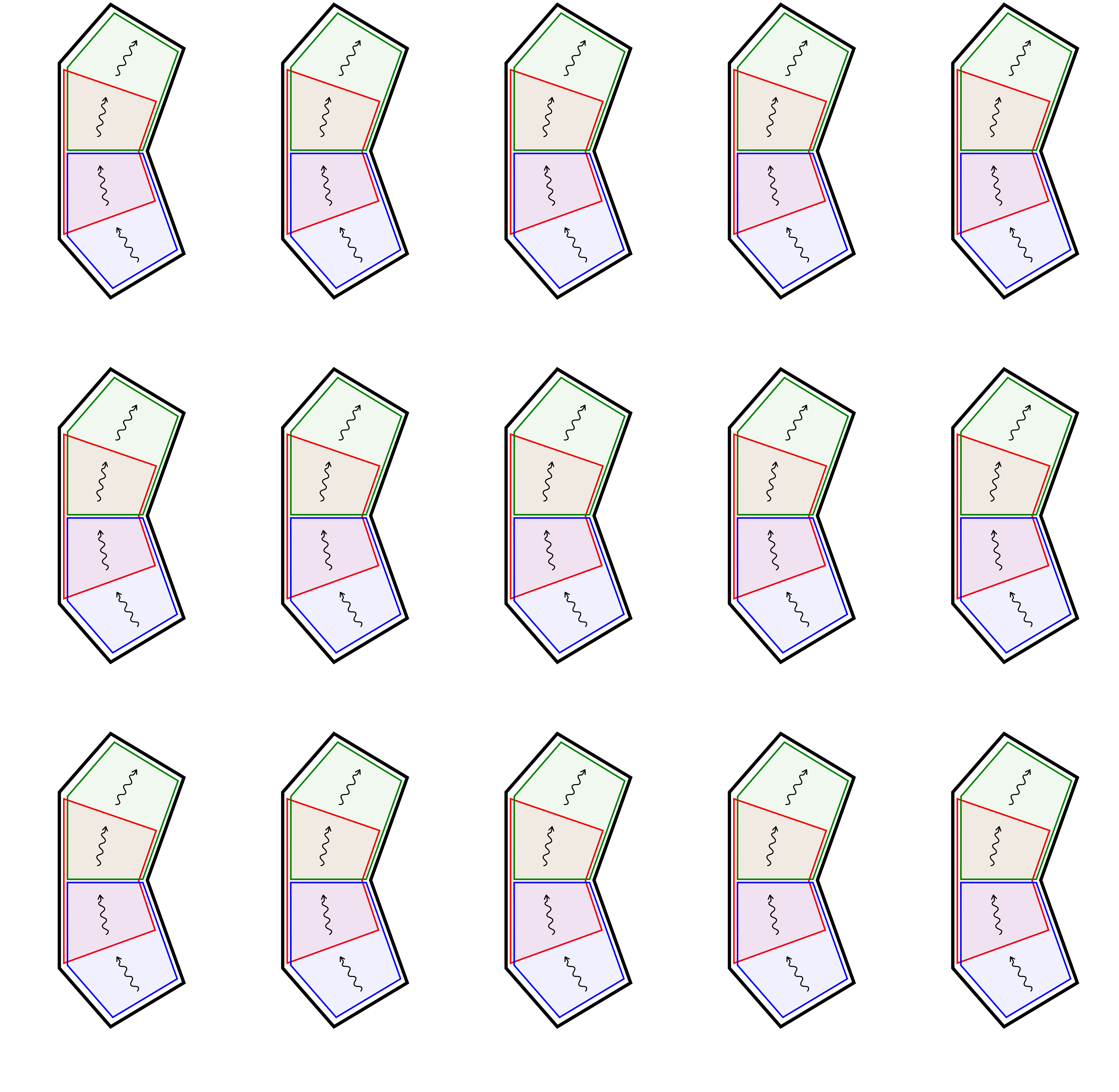
\caption{Fifteen independent components for the NMHV heptagon with the corresponding excitations at twist one. The components in the first line involve charging the bottom and top pentagons only while the second and third line correspond to the remaining ten components where the middle pentagon is also charged. As illustrated here, each such component can be used as a direct probe of a pentagon transition.}
\label{fifteenHep}	
\end{figure}

The heptagon has fifteen independent components \cite{shortSuper} represented in figure \ref{fifteenHep}. Five of them, in the top line of that figure, can be constructed in a similar manner to the hexagon by charging the outermost pentagons.
To generate the remaining ten independent components, it is unavoidable charging the middle pentagon. According to \cite{shortSuper}, charging the middle pentagon typically involves rather nontrivial linear combinations of the super amplitude $\eta$- components. Since we are also interested in testing this map, in what follows we will focus mostly on such examples in which the middle pentagon is charged. 

To leading twist each of the fifteen heptagon components probes a different pentagon transition, see figure \ref{fifteenHep}. We verified that their near collinear expansions are indeed in perfect agreement with the proposals of the previous section once expanded out to leading order in perturbation theory. Interesting as they are, the analysis of these cases follow \cite{data} almost verbatim and is therefore not particularly illuminating to present it in detail. 
Instead, in this section we will consider a richer example involving multi-particle states in both middle squares. A second example can be found in Appendix \ref{anotherEx}. These examples allow one to get a good picture of how generic transitions show up at weak coupling. They also probe considerably more structures in a very non-trivial way and allow us to stress the important role of the so-called \textit{small fermions}. 
The matrix part in the first one is trivially equal to $1$ and in the second case it is simple and has been determined before in \cite{2pt}. 

The examples where more complicated matrices appear were also tested but we leave them for a future publication, where the general construction of the matrix part will be presented. All in all, we have tested \textit{all} possible transitions at tree level up to twists three in one square and two in the second one and several twist three (in both squares) transitions with simple matrix part (in a total of 429 processes).

It is instructive to present one such example in detail. We will analyse the $\cP\circ\cP_{123}\circ\cP_4$ component through the POPE lens. As mentioned above, to make things more interesting and nontrivial we will look at some high twist contribution involving several particles and/or bound-states. To be precise we shall consider the term proportional to 
\beq
e^{-3\tau_1 - 3 i \phi_1} \times e^{-3\tau_2 + 5/2 i \phi_2}  \la{propto}
\eeq
as illustration. What flux tube physical processes govern this contribution? To answer this question it suffices to list everything that has the right quantum numbers to be allowed to flow. In the case at hand we are looking at states with twist $3$ both in the first middle square and the second square. We are searching for states with helicity $-3$ in the first square and $+5/2$ in the second square. Finally, we have $R$-charge considerations. For the sequence $\cP\circ\cP_{123}\circ\cP_4$ we necessarily have an $R$-charge singlet in the first middle square and a state in the fundamental ($\bold{4}$) representation of $SU(4)$ in the second middle square. All in all, this information restricts the matter content enormously. 
\begin{figure}[t]
\centering
\def\svgwidth{16cm} 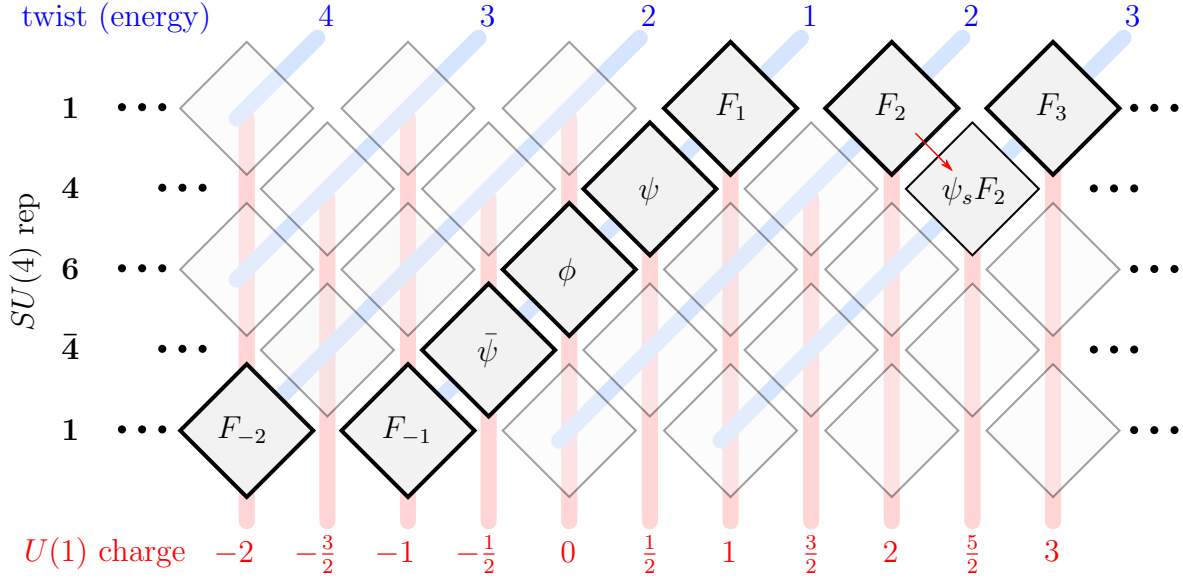
\caption{Fundamental excitations together with the effective excitation $\psi_s F_2\,(\textrm{a.k.a.~} \cD_{1\dot2}^2\psi)$ used in the example. }
\label{TABLEhep}
\end{figure}\\

In the first square, for example, the absolute value of the helicity is maximal, equal to the twist of the state. This saturation is achieved for purely gluonic states only, see figure~\ref{TABLEhep}. There are therefore only three possible states in the first middle square, 
\beq
|F_{-1}(u)F_{-1}(v)F_{-1}(w)\>\,,\qquad |F_{-2}(u)F_{-1}(v)\>\,,\qquad |F_{-3}(u)\>\,.
\eeq
The first two are multi-particle states and kick in at higher loop orders. The last one, corresponding to a bound-state of three negative helicity gluons, is the only one showing up at leading order at weak coupling. 

In the second square things are more interesting. With helicity $5/2$ there are only two possible states we could envisage:
\beq
|F_{1}(u)F_{1}(v)\psi(w)\>\,,\qquad |\psi(w)F_{2}(u)\>\,. \la{list2}
\eeq
At first we could imagine discarding both since they are both multi-particle states; however, when fermions are involved the coupling analysis is more subtle. The point is that fermions can be either small or large and in the former case they act as sort of symmetry generators~\cite{2pt}. As such each of the states in (\ref{list2}) can be split into two cases depending on whether the fermion is small or large. In particular, 
the second state $|F_{2}(u)\psi(v)\>$ with the fermion evaluated in the small fermion domain can be seen as a supersymmetry generator acting on the excitation $F_{2}(u)$ thus generating a \textit{single} effective weak coupling excitation -- see figure~\ref{TABLEhep} -- and as such might show up already at leading order at weak coupling. (At the same time the first state $|F_{1}(u)F_{1}(v)\psi(w)\>$ with $\psi$ being a small fermion would behave as a two particle state and thus show up only at higher orders in perturbation theory.)

In sum, to match against tree level data it suffices to focus on the process
\beq
\texttt{vacuum} \to F_{-3}(u) \to\psi(w)  F_{2}(v)\to \texttt{vacuum}\,,
\eeq
which is what we turn to now. In this case the matrix part is trivial. Indeed, the R-charge index of the fermion is unambiguously fixed once we pick an R-charge configuration for the various $\chi$'s. In other words, the matrix part in (\ref{POPEintegrand}) is equal to one and the full integrand is just a product of the dynamical part and the form factor contribution. According to (\ref{dynP}) and~(\ref{ffactors}) these read
\beqa
\texttt{dynamical part}\!\!&=&\!\!  \hat\mu_{F_{-3}}(u)\,\hat\mu_{\psi}(w)\,\hat\mu_{F_{2}}(v)\,\frac{P_{\bar\psi|F_{3}}(w|u)P_{F_{-2}|F_{3}}(v|u)}{P_{\bar\psi|F_{-2}}(w|v)P_{F_{2}|\psi}(v|w)}\,,\\
\texttt{form factors part}\!\!&=&\!\! \frac{1}{{g^{\frac{3}{4}}}}\, (h_{F_{3}}(u))^0\, (h_{F_{-3}}(u))^3\, (h_{\bar\psi}(w))^3\,  (h_{F_{-2}}(v))^3\, h_{\psi}(w)\, h_{F_{2}}(v) \,,
\eeqa
which we simply multiply together to obtain the {POPE} prediction
\beq
\mathcal{W}_{F_{-3}\rightarrow\psi \, F_2}=\int\limits_{\mathbb{R}} \frac{du}{2\pi} \int\limits_{\mathbb{R}} \frac{dv}{2\pi} \int\limits_{ {\bar{\mathcal{C}}_\text{small}}}\frac{dw}{2\pi} \,\, (\texttt{dynamical part} )\times (\texttt{form factors part})\, ,
\eeq
{where $\bar{\mathcal{C}}_\text{small}$ is the \textit{complex conjugate} version of the half-moon contour mentioned below~(\ref{hexagons}), hence running in the upper half $u$ plane, in agreement with the alternating conventions in~(\ref{dynP}). (Equivalently, depending on the $\pm$ sign in front of $ip\sigma$ in~(\ref{dynP}) we use an $\pm i\epsilon$ prescription for integration around zero momentum fermions in the corresponding square.)}
Plugging all the building blocks together, we therefore arrive at 
\beqa
\mathcal{W}_{F_{-3}\rightarrow\psi \, F_2}&=&e^{-3\tau_1 - 3 i \phi_1} \times e^{-3\tau_2 + 5/2 i \phi_2} \times \\
&&\!\!\!\!\!\!\!\!\!\!\!\!\!\!\!\!\!\!\!\!\!\!\!\!\!\!\!\!\times \int\limits_{\mathbb{R}} \frac{du}{2\pi} \int\limits_{\mathbb{R}} \frac{dv}{2\pi} \int\limits_{{\bar{\mathcal{C}}_\text{small}}}\frac{dw}{2\pi}\,\frac{w\, \Gamma \left(\frac{5}{2}+i u\right) \Gamma (2-i v) \Gamma \left(\frac{7}{2}-i u+i v\right)}{2i\left(u^2+\frac{9}{4}\right) \left(v^2+1\right) \left((v-w)^2+1\right)} e^{2i u \sigma_1 -2 i (v+w) \sigma_2}+\cO(g^2)\,, \nn
\eeqa
{and verify that, despite the funny fractional powers of the coupling appearing in the individual ingredients, the resulting integrand has a regular expansion in $g^2$ and starts at tree level, as expected. (This phenomenon is not accidental and is discussed further in {the conclusions} of the paper.)}

One of the three integrals involves a small fermion $\psi(w)$ integrated over its corresponding small fermion contour $ {\bar{\mathcal{C}}_\text{small}}$. An important universal property of small fermions is that they can always be straightforwardly integrated out (at any value of the coupling in fact). In this tree level example we see that the only singularity inside the half moon encircling the  {upper} half plane is the single pole at $ {w=v+i}$. The fermion integral thus collapses into the corresponding residue contribution which freezes $w$ to be attached to the rapidity $v$ in a Bethe string like pattern. The interpretation of such strings is that the fermion is acting as a symmetry generator on the other excitation in this square, the bound-state of gluons $F_2(v)$. The result of this action is an effective twist $3$ weak coupling excitation, see figure \ref{TABLEhep}, which is described by the Bethe string. 

In sum, after integrating out the small fermion we end up with the integrations in $u$ and $v$ for a single effective particle in each square. The resulting integral can then be straightforwardly performed leading to the prediction
\beq
\!\!\!\! \begin{aligned} \label{result}
\mathcal{W}_{F_{-3}\rightarrow\psi \, F_2}&=\frac{e^{-3\tau_1-3\tau_2- 3 i\phi_1+ 5 i\phi_2/2}}{\left(e^{2 \sigma _1}+1\right)^3 \left(e^{2 \sigma _2}+1\right)^3 \left(e^{2 \sigma _2+2\sigma_1} +e^{2 \sigma _2}+e^{2 \sigma _1}\right)^5}\times(e^{13 \sigma _1+2 \sigma _2}+5\, e^{11 \sigma _1+4 \sigma _2}+8\, e^{13 \sigma _1+4 \sigma _2}\\
&\!\!\!\!\!\!+10 \,e^{9 \sigma _1+6 \sigma _2}+35\, e^{11 \sigma _1+6 \sigma _2}+28\, e^{13 \sigma _1+6 \sigma _2}+10\, e^{7 \sigma _1+8 \sigma _2}+60\, e^{9 \sigma _1+8 \sigma _2}+105\, e^{11 \sigma _1+8 \sigma _2}\\
&\!\!\!\!\!\!+56 \,e^{13 \sigma _1+8 \sigma _2}+5 \,e^{5 \sigma _1+10 \sigma _2}+35 \,e^{7 \sigma _1+10 \sigma _2}+105\, e^{9 \sigma _1+10 \sigma _2}+130 \,e^{11 \sigma _1+10 \sigma _2}+55\, e^{13 \sigma _1+10 \sigma _2}\\
&\!\!\!\!\!\!+e^{3 \sigma _1+12 \sigma _2}+8\, e^{5 \sigma _1+12 \sigma _2}+28 \,e^{7 \sigma _1+12 \sigma _2}+56\, e^{9 \sigma _1+12 \sigma _2}+55 \,e^{11 \sigma _1+12 \sigma _2}+20 \,e^{13 \sigma _1+12 \sigma _2})\,.
\end{aligned}
\eeq

This example clearly illustrates the importance of checking the integrability against perturbative data. After all, it  is clearly a tall order to reproduce any result of the complexity of (\ref{result}). According to the proposal in \cite{shortSuper}, the amplitude $\cP\circ\cP_{123}\circ\cP_4$  can be extracted from standard $\eta$-components as 
\beqa
\Big(\frac{\partial}{\partial\chi_2}\Big)^3\!\!\frac{\partial}{\partial\chi_3}\mathcal{W}\!\!\!\!&=&\!\!\!\!\dfrac{(\textbf{5})_3}{\((\textbf{1})_2(\textbf{2})_2(\textbf{3})_2\)^3}\(\<1,2,3,-1\>\dfrac{\partial}{\partial\eta_{-1}}+\<1,2,3,0\>\dfrac{\partial}{\partial\eta_{0}}\)^3 \dfrac{\partial}{\partial\eta_5} \mathcal{W}\\
\!\!\!\!&=&\!\!\!\!\dfrac{(\textbf{5})_3}{\((\textbf{1})_2(\textbf{2})_2(\textbf{3})_2\)^3} \(\<1,2,3,-1\>^3\,\mathcal{W}^{(-1,-1,-1,5)}+\<1,2,3,-1\>^2\<1,2,3,0\>\,\mathcal{W}^{(-1,-1,0,5)}\right. \nonumber\\
&&\left.+\,\<1,2,3,-1\>\<1,2,3,0\>^2\,\mathcal{W}^{(-1,0,0,5)}+\<1,2,3,0\>^3\,\mathcal{W}^{(0,0,0,5)}\) \label{linear}\,.
\eeqa 
Each of these components can be obtained by running similar code lines in Mathematica as in the previous example of the hexagon, using the heptagon twistors in the Appendix A of~\cite{data}. Once we evaluate the brackets and the weights, we expand the result at large $\tau_1$ and $\tau_2$ and pick the term proportional to (\ref{propto}). In this way we obtain a perfect match with the expression (\ref{result})! 

These are formidable checks of the full POPE construction as they are probing, at the same time, the map between charging pentagons and charging edges of \cite{shortSuper} as well as the (weak coupling expansion of) the various elements of the POPE integrand. We performed several other checks of this sort (more than a hundred of them) always obtaining a perfect match. For completeness, we present another example in appendix \ref{anotherEx}. We also explored some higher loop data but our analysis there was much less thorough. It would be interesting to push it much further both in higher twists and higher loops. {In particular, it would be nice to make contact with the very interesting recently uncovered heptagon bootstrap \cite{heptagonB}.}

\section{Concluding Remarks}
In this paper we put forward the full coupling dependent part of the POPE integrand. This is an important step towards the completion of the OPE program stated in the beginning. 

One of the most amusing features of the POPE which renders it so efficient at weak coupling is that the number of particles contributing at a given loop order grows very slowly with the loop order. 

\begin{figure}[t]
\centering
\def\svgwidth{16cm} 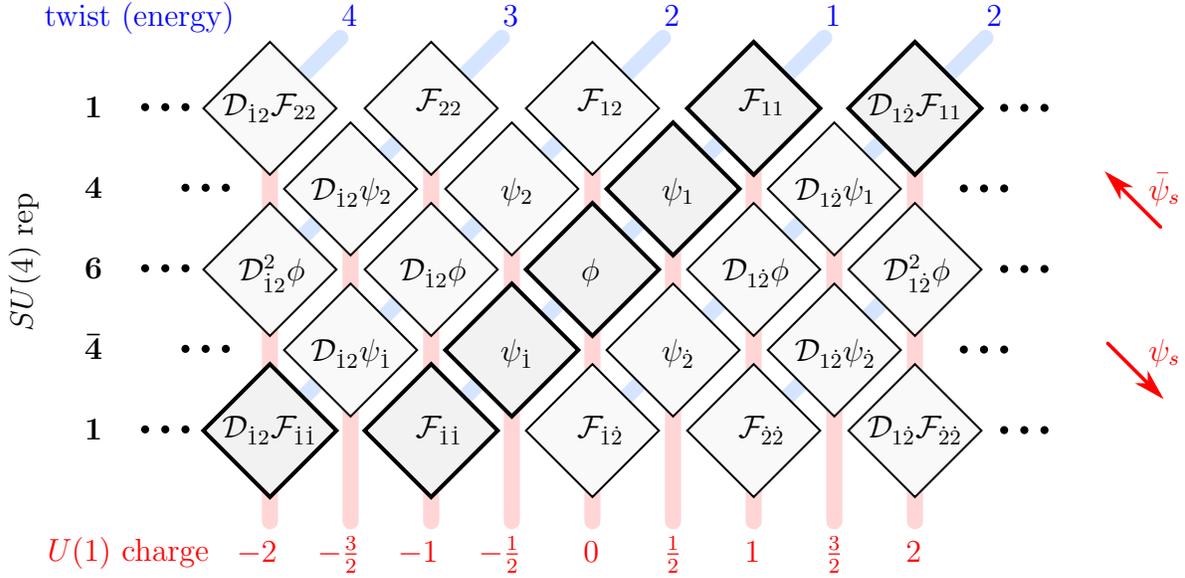
\caption{Table of weak coupling excitations of the flux tube. The fundamental excitations are sitting on the boldfaced squares. Additionally one can generate a plethora of effective particles, by attaching small fermions $\psi_s$ or $\bar{\psi}_s$  to these fundamental excitations, as they act as supersymmetry generators. Besides the excitations represented in this table, one could generate other excitations by attaching both $\psi_s$ and $\bar{\psi}_s$ to a fundamental excitation.  This gives rise to a third dimension not represented here, where excitations involving $\mathcal{D}_{2\dot{2}}$ are sitting.}
\label{excitationsTable}
\end{figure}

To be precise we should recall the notion of weak coupling \textit{effective particles} and \textit{small fermions}. What happens at weak coupling is that on top of any number of fundamental excitations -- that is gluons and their bound-states, scalars and fermions -- we can add arbitrarily many so-called small fermions $\psi$ or $\bar\psi$ which behave roughly as supersymmetry generators acting on the fundamental excitations and morphing them into what one calls {\it effective particles}, see figure \ref{excitationsTable}. Technically the way this works is that the POPE integrals over the small fermions can always be integrated out explicitly leading to string-like structures attached to the rapidities of the fundamental excitations. These strings are the mathematical depiction of the effective particles. 

To estimate the coupling dependence of each POPE process we simply take the POPE integrand (\ref{OPEintegrand}) for a given matter content and look for the leading power of $g$ of each of the factors in the dynamical and form factor part (the matrix part being coupling independent). When doing this we should pay special attention to the fermions as small and large fermion measures, transitions and form factors scale quite differently at weak coupling. At the end of the day, the results we find are {remarkably simple}. Take the hexagon for instance. We readily find that each OPE process scales as 
\beq
g^{\,M^2+\(K-\frac{r_b-r_t}{4}\)^2-\frac{1}{16}(r_t+r_b)(8-r_t-r_b)} \la{estimate} \,,
\eeq
where $r_b$ are the number of $\chi$'s in the bottom pentagon and $r_t$ is the number of $\chi$'s of the top pentagon. (The fact that $r_t=r_b=0$ is equivalent to $r_t=r_b=4$ is the statement that for 6 points MHV and N$^2$MHV are trivially related by parity to one another. For NMHV we have $r_t=4-r_b$.) Finally $K$ and $M$ contain the information about the matter content. We have
\beq
M=N_{F}+N_{\psi}+N_\phi+N_{\bar \psi}+N_{\bar F} \qquad \text{and} \qquad K= N_{F}+\frac{1}{2}N_{\psi}-\frac{1}{2}N_{\bar \psi}-N_{\bar F}-\frac{1}{2}N_{\psi_S}+\frac{1}{2}N_{\bar \psi_S} \,. \la{MK}
\eeq
Here $N_{F}$ ($N_{\bar F}$) indicate the total number of gluons of positive (negative) helicity or their bound-states and $N_{\psi}$ denotes the number of usual (i.e.~large) fermions while $N_{\psi_S}$ is the number of small fermions. The exponent in (\ref{estimate}) is always an even number (which is nothing but twice the number of loops).
Let us expand a bit on the physics of (\ref{estimate}) and (\ref{MK}).

We see that $M$ is the total number of excitations except for small fermions. In other words, since the small fermions can always be integrated out to simply change the flavour of the other excitations, $M$ is nothing but the total number of effective excitations, 
\beq
M=\texttt{number of effective excitations}\,. \la{intM}
\eeq
Take for example a process with $6$ effective excitations for the MHV hexagon. From (\ref{estimate}) we see that it will first show up at eighteen loops (if $K=0$) or even later (if $K\neq 0$). Up to seven loops, for instance, we shall never need more than $3$ effective particles to describe the six point MHV scattering amplitude! 

Next we have $K$ which contrary to $M$ can be positive or negative. Consider first a configuration containing only fundamental excitations and no small fermions. Then $K$ is the average of ratios of the $U(1)$ charge to bare twist of each excitation. It is a sort of measure of helicity violation. Equivalently, if we associate a weight of $1-j/2$ to an excitation in the $j$-th row in figure \ref{excitationsTable} then $K$ is the average of these weights which we can therefore depict as a vertical centre of mass position of sorts in this figure. This second description is the most convenient one to generalize to the case where we add small fermions to the mix. As indicated by the red arrows in figure \ref{excitationsTable}, such small fermions transform the various excitations moving them up or down in this table. With the precise signs in (\ref{MK}) we see that the interpretation as the average of rows is perfectly kept when we move to a description purely in terms of effective excitations. In short, 
\beq
K=\texttt{vertical centre of mass of figure \ref{excitationsTable}}\,. \la{intK}
\eeq

In sum, OPE processes are coupling suppressed by two effects: large number of particles and large helicity violations. 

We see that for a given number of effective particles $M$ the processes that minimize the loop order are those for which $K=\frac{r_b-r_t}{4}$. For the NMHV hexagon this means $K=\frac{r_b-2}{2}$ which translates into the intuitive statement that the centre of mass position $K$ should coincide with that of the leading order excitation flowing in that hexagon as identified in figure \ref{FIGhexagon}{b}. Conversely, for a given number of effective particles $M$ the processes that maximize the loop order are those for which $K$ is as large as possible (positive or negative depending on weather $r_b$ is less or greater than $r_t$). For MHV, for instance, the processes that maximize $|K|$ are the states containing nothing but gluons. In that case (\ref{estimate}) reduces to
\beq
g^{2N_F^2+2N_{\bar F}^2}\label{loop-c}
\eeq
reproducing the counting in \cite{fusion}, see equation (44) there. As explained in \cite{fusion}, because these states maximize $U(1)$ charge they are very easily identified in perturbation theory by taking a so-called double scaling limit. {This sector also appears to be instrumental in making contact with the hexagon bootstrap program~\cite{LanceEtAl1,LanceEtAl2,LanceEtAl3,LanceEtAl4,LanceEtAl5} as thoroughly explored recently in~\cite{georgiosjames}, using technology from~\cite{georgios}. In this regard, it would be interesting to see if the OPE loop thresholds~(\ref{loop-c}) parallel some sort of complexity jumps or qualitative changes in the hexagon function representations.}

A similar {loop counting} can be straightforwardly performed for higher $n$-gons where we find that a given OPE sequence first shows up at $g^{2l}$ with the number of loops $l$ given by
\beq
2l=\sum\limits_{i=1}^{n-5}\[M_i^2-M_i M_{i+1}+\(K_i-\frac{r_i-r_{i+1}}{4}\)^2-K_i K_{i+1}\]-\sum\limits_{i=1}^{n-4}\frac{1}{16}\,r_i\,(8-r_{i-1}-r_{i+1}) \,.
\eeq
Here $M_i$ and $K_i$ are associated to the $i$-th middle square and are defined, for a given matter content flowing in that square, exactly as above (\ref{MK}).\footnote{{The values for $M_{n-4}$ and $K_{n-4}$ should be set to zero.}} Obviously, the interpretation as (\ref{intM}) and (\ref{intK}) continues to hold. Similarly, the $r_i$ are the number of $\chi$'s in the $i$-th pentagon with $r_{0}\equiv r_1$ for $i=1$ and $r_{n-3} \equiv r_{n-4}$ for the boundary cases. Note in particular that for heptagons and highers the loop counting is not simply a sum of squares. Now some terms contribute with a minus sign. An interesting outcome of this fact is that with higher $n$-gons we can often engineer processes with large number of particles at relatively low loop order by considering polygons with many edges and slowly injecting more and more particles and $\chi$'s as we move along the tessellation, in the same way as one efficiently starts a car by gently pushing it down a road. The longer the road, and the more help one gets along the way, the easiest it is to start the car. This is often quite useful if one wishes to test higher particle contributions at weak coupling without going to prohibitively high loop orders. Some examples of this enhancement were already discussed in a purely gluonic context in \cite{fusion}. For instance, we can set all $r$'s to two, start with a scalar in the first square and add one scalar in each square until a maximum halfway through the tessellation where we start decreasing one by one the number of scalars. A simple counting exercise shows that this configuration will first appear at tree-level for any polygon. To give an example, we could check a process involving 4998 scalars in a myriagon already at tree-level!

In \cite{FrankToAppear} the full matrix part for any $n$-gon, the last missing piece in the POPE integrand, is analysed and in \cite{shortHexagon} the outcome of this analysis shall be unveiled for the hexagon case. Together with the result herein presented these works flash out the initial proposal in \cite{short} to completion. Nonetheless, in our view, they are not  simply the end. Having a fully non-perturbative proposal for scattering amplitudes in a four dimensional gauge theory is very exciting but more exciting still is the perspective of using it to extract sharp physics out of it at weak, strong and finite coupling, thus substantially enlarging our understanding of non-perturbative quantum field theory. 

\section*{Acknowledgements} 
We are obliged to A.~Belitsky, J.~Bourjaily, S.~Caron-Huot, F.~Coronado and J.~Maldacena for inspiring discussions. We also acknowledge all the participants of the `Program on Integrability, Holography and the Conformal Bootstrap' for an inspiring and exciting program. We would like to thank FAPESP grant 2011/11973-4 for funding our visit from Nov 2014 to March 2015 to ICTP-SAIFR where most of this work was done. Research at the Perimeter Institute is supported in part by the Government of Canada through NSERC and by the Province of Ontario through MRI. L.C. is funded by a CONACyT doctoral scholarship. A.S. has been supported by the I-CORE Program of the Planning and Budgeting Committee, The Israel Science Foundation (grant No. 1937/12) and the EU-FP7 Marie Curie, CIG fellowship. J.C. is funded by the FCT fellowship SFRH/BD/69084/2010. The research leading to these results has received funding from the People Programme (Marie Curie Actions) of the European Union's Seventh Framework Programme FP7/2007-2013/ under REA Grant Agreement No 317089 (GATIS). Centro de Fisica do Porto is partially funded by the Foundation for Science and Technology of Portugal (FCT). The research leading to these results has received funding from the [European Union] Seventh Framework Programme [FP7-People-2010-IRSES] under grant agreements No 269217.

\appendix

\section{Pentagon transitions and measures}\label{alltransitions}

In this appendix we summarize our knowledge about elementary transitions $P_{X|Y}$, with $(X,Y)$ being any pair of flux tube excitations. Their general structure is
\beq
\begin{aligned}\la{FandExponent}
P_{X|Y}(u|v) &= f_{X}(u)F_{XY}(u,v)f_{Y}(-v)\\
&\times\exp\Big[2(\kappa_X(u)-i\tilde\kappa_X(u))^{t}\cdot \mathcal{M}\cdot\kappa_Y(v)+2i(\kappa_X(u)+i\tilde\kappa_X(u))^t\cdot \mathcal{M}\cdot\tilde\kappa_Y(v)\Big] \, ,
\end{aligned}
\eeq
where all the objects in the exponent were explicitly given in the appendix C of \cite{2pt} for all sorts of excitations of the flux tube (with $\mathcal{M} = \mathbb{Q} \cdot\mathbb{M}$ in the notations of~\cite{2pt}, see also~\cite{fusion}). We also found convenient to strip out the factor
\beq\label{f-fact}
\log f_{X}(u) = \int\limits_{0}^{\infty}\frac{dt}{t}(J_{0}(2gt)-1)\frac{\frac{1}{2}J_{0}(2gt)+\frac{1}{2}-e^{-q_Xt}e^{-iut}}{e^{t}-1}\, ,
\eeq
for each excitation, with $J_0(z) = 1+O(z^2)$ the Bessel function of the first kind and $q_{X} = -1/2, 0, 1/2, 1, \ldots$ for scalar, large fermion, elementary gluon, bound state of two gluons, etc. (Note that in the case of a small fermion, i.e.~$X = \psi_{S}$ or $\bar{\psi}_{S}$, we have $f_{X}(u) = 1$ identically, see~\ref{APPsmall} below.)

The factor~(\ref{f-fact}) as well as the term in the exponent above are quite universal and, in particular, only depend on the absolute values of the $U(1)$ charges (e.g.~they cannot distinguish between $(X, Y) = (\psi, \psi), (\psi, \bar{\psi}), (\bar{\psi}, \psi),$ or $(\bar{\psi}, \bar{\psi})$). As such, the function $F_{XY}(u, v)$ has the same conjugation property as its parent transition. Since all our transitions obey%
\footnote{We failed to find a reason for this simple property, but noticed that it is consistent both with the fundamental relation~(\ref{fundamentalrelation}), since $S_{XY}(u, v)^* = S_{YX}(v, u) = S_{\bar{Y}\bar{X}}(v, u)$, and with the mirror equation~(\ref{mirror}). In the latter case, one needs to use that $u^{-\gamma}$ turns into $u^{+\gamma}$ upon conjugation and that~(\ref{mirror}) is equivalent to $P_{X|Y}(u|v^{\gamma}) = P_{Y|\bar{X}}(v|u)$.}
\beq\label{conj}
P_{X|Y}(u|v)^* = P_{\bar{Y}|\bar{X}}(v|u)\, ,
\eeq
upon complex conjugation (for real rapidities), with of course $\bar{\phi} = \phi$ for a scalar, $\bar{F}_{a} = F_{-a}$ for a gluon, etc., then the exact same relation holds true for the corresponding functions $F$.

It is also interesting to note that both $\log f_{X}$ and the exponent in~(\ref{FandExponent}) are of order $O(g^2)$ for small $g$. (This estimate is not uniform in the rapidities and holds only away from the locations of singularities, which are at imaginary half integer values at weak coupling.) The leading order weak coupling results can thus be directly obtained from the prefactors $F_{XY}$.

\subsection{Summary of transitions}\label{sump}

Knowing the transitions is equivalent to knowing the prefactors $F$ in~(\ref{FandExponent}). For them, which as we just said are also all we need to know to leading order at weak coupling, we have the following lists. {(Up to few exceptions involving gluonic bound states, all the transitions given below already appeared in~\cite{data,2pt,Andrei1,fusion,Andrei2,Belitsky:2014sla}.%
\footnote{Mixed transitions involving gluonic bound states were independently obtained by A.~Belitsky~\cite{Andrei-comm}.}
At the end of this subsection, we comment on the `difference of normalizations' between our transitions here and those obtained in these series of papers.)}

\subsubsection*{Transitions involving a gluon or a bound state of gluons}

We start by the cases involving a gluon or a bound state of gluons.

The purely gluonic transitions are given by 
\begin{eqnarray}
\begin{aligned}\label{Pg1}
&{F}_{F_aF_b}(u,v)&=&\quad\sqrt{(x^{[+a]}y^{[-b]}-g^2)(x^{[-a]}y^{[+b]}-g^2)(x^{[+a]}y^{[+b]}-g^2)(x^{[-a]}y^{[-b]}-g^2)}  \\
&&&  \times\,\frac{(-1)^b \Gamma(\frac{|a|-|b|}{2}+iu-iv) \Gamma(\frac{|a|+|b|}{2}-iu+iv)}{g^2 \Gamma(1+\frac{|a|}{2}+iu) \Gamma(1+\frac{|b|}{2}-iv) \Gamma(1+\frac{|a|-|b|}{2}-iu+iv)} \, , \,\,\,\,\,\,\, \text{for }ab> 0\,, \noindent\\
&{F}_{F_aF_{b}}(u,v)&=&\quad\frac{1}{\sqrt{(1-\frac{g^2}{x^{[+a]} y^{[-b]}})(1-\frac{g^2}{x^{[-a]} y^{[+b]}})(1-\frac{g^2}{x^{[+a]} y^{[+b]}})(1-\frac{g^2}{x^{[-a]} y^{[-b]}})}} \\
&&& \times\,\frac{\Gamma(1+\frac{|a|+|b|}{2}+iu-iv)}{\Gamma(1+\frac{|a|}{2}+iu) \Gamma(1+\frac{|b|}{2}-iv)} \, , \,\,\,\,\,\,\, \text{for }ab<0\,,\noindent \\
\end{aligned}
\end{eqnarray}
and the mixed ones by
\begin{equation}\label{Pg2}
{F}_{F_a \phi }(u,v) = {F}_{\phi F_a}(-v,-u)= \frac{ \sqrt{x^{[+a]}x^{[-a]}}\Gamma(\frac{1}{2}+\frac{\left|a\right|}{2}+iu-iv)}{g \Gamma(1+\frac{|a|}{2}+iu)  \Gamma(\frac{1}{2}-iv)} \, ,
\end{equation}
for any $a$, and by
\begin{equation}
\begin{aligned}\label{Pg3}
&{F}_{F_a\psi}(u,v) = {F}_{\psi F_a}(-v,-u) = -\frac{iy(x^{[+a]}x^{[-a]})^{3/4}\Gamma(\frac{a}{2}+iu-iv) \sqrt{(1-\frac{g^2}{x^{[+a]}y})(1-\frac{g^2}{x^{[-a]}y})}}{g^{3/2}\Gamma(1+\frac{a}{2}+iu)\Gamma(1-iv)}\, , \\
&{F}_{F_{-a}\psi}(u,v) = {F}_{\psi F_{-a}}(-v,-u)  =\frac{\,  {(x^{[+a]}x^{[-a]})^{1/4}}\,\Gamma(1+\frac{a}{2}+iu-iv)}{g^{1/2} \Gamma(1+\frac{a}{2}+iu)\Gamma(1-iv) \sqrt{(1-\frac{g^2}{x^{[+a]}y})(1-\frac{g^2}{x^{[-a]}y})}}\, ,
\end{aligned}
\end{equation} 
for $a > 0$. In all cases, we have $x=x(u),\, x^{[\pm a]}=x(u\pm ia/2),\, y=x(v),\, y^{[\pm b]}=x(v\pm ib/2), x(u) \equiv \frac{1}{2}(u+\sqrt{u^2-(2g)^2})$ and $\Gamma(z)$ the Euler Gamma function. Transitions involving $\bar{\psi}$ can be obtained by conjugating those with $\psi$, as in~(\ref{conj}).

(Note that all the factors above are normalized such that the associated transitions are equal to $1$ to leading order at strong coupling, in the perturbative regime, i.e.~for excitations with momenta of the same order as their masses. This is the expected decoupling property of the gluons at strong coupling.)

\subsubsection*{Transitions involving only scalars or fermions }

We proceed with the remaining set of functions $F_{XY}$ involving scalars and fermions. They read
\beq
\begin{aligned}\label{Psf}
&{F}_{\phi\phi}(u,v)=\frac{\Gamma(i u-i v)}{g\Gamma(\frac{1}{2}+iu) \Gamma(\frac{1}{2}-iv)} \,,\\
&{F}_{\phi \psi} (u,v) = {F}_{\phi \bar{\psi}} (u,v) = \frac{\sqrt{y}\,\Gamma \left(\frac{1}{2}+iu-iv\right)}{g\Gamma(\frac{1}{2}+iu)\Gamma(1-iv)}\, ,\\
&{F}_{\bar{\psi}\psi}(u,v) = -{F}_{\psi \bar{\psi}}(u,v)  = \frac{(xy)^{3/4}\Gamma (1+iu-iv)}{g^{3/4}\Gamma (1+iu)\Gamma (1-iv) \sqrt{xy-g^2}}\,, \\
&{F}_{\psi\psi}(u,v) = -{F}_{\bar{\psi}\bar{\psi}}(u,v) = \frac{i(x y)^{1/4}\Gamma(iu-iv) \sqrt{xy-g^2}}{g^{5/4}\Gamma (1+iu) \Gamma (1-iv)}\,.
\end{aligned}
\eeq
{(The branch choice for the mixed transitions above was mostly driven by the goal of getting the sign-free large (positive) $v$ behaviours~(\ref{zerof}).)}

The lists~(\ref{Pg1}),~(\ref{Pg2}),~(\ref{Pg3}) and~(\ref{Psf}) cover all the pentagon transitions of the OPE program.

\subsubsection*{{Comparison with the literature}}

{As alluded to before, the pentagon transitions listed before already appeared in the literature. We found useful however to redefine some of them, still preserving the fundamental relation~(\ref{fundamentalrelation}) and the mirror axiom~(\ref{mirror}). For instance, comparing the scalar transition in~(\ref{Psf}) with the one appearing in~\cite{data}, we find that}
\beq
P_{\phi|\phi}(u|v)_{\textrm{here}} = gP_{\phi|\phi}(u|v)_{\textrm{\cite{data}}}\, .
\eeq
{Clearly, such an innocent rescaling cannot alter the validity of~(\ref{fundamentalrelation}) and~(\ref{mirror}). A slightly more involved redefinition is found when comparing our mixed transitions in~(\ref{Pg3}) with those found in~\cite{Andrei1}. We get}
\beq
\begin{aligned}
&P_{F|\psi}(u|v)^2_{\textrm{here}} = -\frac{\sqrt{x^{+}x^{-}}}{g}\times P_{F|\psi}(u|v)^2_{\textrm{\cite{Andrei1}}}\, , \\
&P_{F|\bar{\psi}}(u|v)^2_{\textrm{here}} = -\frac{g}{\sqrt{x^{+}x^{-}}}\times P_{F|\bar{\psi}}(u|v)^2_{\textrm{\cite{Andrei1}}}\, .
\end{aligned}
\eeq
{Nonetheless, here as well, both expressions fulfill the same axioms~(\ref{fundamentalrelation}) and~(\ref{mirror}), and thus differ only by CDD factors. Along a somewhat similar vein,}%
\footnote{{In case of fermions, it is much harder to check which choice is better as far as the mirror property is concerned, since the mirror algebra is anomalous for fermions and thus harder to implement, see appendices of~\cite{data} for more details.}}
{we have}
\beq
\begin{aligned}
P_{\psi|\psi}(u|v)_{\textrm{here}} = i\frac{g^{3/4}}{(xy)^{1/4}}\times P_{\psi|\psi}(u|v)_{\textrm{\cite{2pt}}}\, , \qquad P_{\psi|\bar{\psi}}(u|v)_{\textrm{here}} = -\frac{(xy)^{1/4}}{g^{3/4}}\times P_{\psi|\bar{\psi}}(u|v)_{\textrm{\cite{2pt}}}\, ,
\end{aligned}
\eeq
{and also}
\beq
P_{\bar{\psi}\bar{\psi}}(u|v)_{\textrm{here}} = -P_{\psi\psi}(u|v)_{\textrm{here}}\, , \qquad P_{\bar{\psi}\psi}(u|v)_{\textrm{here}} = -P_{\psi\bar{\psi}}(u|v)_{\textrm{here}}\, ,
\eeq
{while the convention of~\cite{2pt} was}
\beq
P_{\bar\psi \bar\psi}(u|v)_{\textrm{\cite{2pt}}} = P_{\psi \psi}(u|v)_{\textrm{\cite{2pt}}} \,, \qquad P_{\psi \bar\psi}(u|v)_{\textrm{\cite{2pt}}} = P_{\bar{\psi} \psi}(u|v)_{\textrm{\cite{2pt}}} \, .
\eeq
{These are all the differences we could find. As we can see, they almost all relate to the choice of CDD factors, which (by definition) is the sole freedom left after imposing the bootstrap axioms.%
\footnote{In the conventions of~\cite{2pt} the fundamental axiom~(\ref{fundamentalrelation}) for $\psi\bar{\psi}$ was not including a minus sign in front of the S-matrix; hence differences in transitions related to this sign are not strictly speaking of CDD type.}
This ambiguity is not without connection to the freedom we have to reshuffle factors in the definitions of the various elements of the POPE integrand, provided we have the same physical integrand once we multiply them together. This is precisely what happens when relating the integrands that we can produce here with those derived in~\cite{data,2pt,Andrei1,Andrei2,Belitsky:2014sla} using slightly different transitions, measures, and, sometimes, rules for assigning form factors. Our conventions here have the slight advantage that no form factors are needed for MHV amplitudes; equivalently, they are normalized such that the relation between charged pentagons and zero momentum fermions is most straightforward. It is nonetheless clearly desirable to find a better handle on the ambiguities mentioned above. Additional constraints should, for instance, follow from the nonlinear nature of the anomalous mirror rotation for fermions and/or from certain fusion properties the various transitions could obey. It remains to work them out and see how much they can tell.}

\subsection{Analytic continuation to small fermions}\la{APPsmall}

It is also convenient to store the representation for small fermions. This one is obtained by direct analytical continuation, $x(v)\rightarrow x(\check{v}) = g^2/x(v)$, and it was thoroughly exemplified in \cite{2pt}. The squared transitions, for instance, read as :
\beq
\begin{aligned}
&P_{\phi\psi}(u|\check{v})^2 =- \frac{S_{\phi\psi}(u, \check{v})}{(u-v+\frac{i}{2})S_{\star \phi\psi}(u, \check{v})}\, ,\\
&P_{F\psi}(u|\check{v})^2 = -\frac{{g }{\sqrt{x^{+}x^{-}}}y(u-v-\frac{i}{2})S_{F\psi}(u, \check{v})}{\,(x^{+}y-g^2)(x^{-}y-g^2)S_{\star F\psi}(u, \check{v})}\, ,\\
&P_{F\bar{\psi}}(u|\check{v})^2 = -\frac{(x^{+}y-g^2)(x^{-}y-g^2)S_{F\bar{\psi}}(u, \check{v})}{{g}\,{\sqrt{x^{+}x^{-}}}y(u-v-\frac{i}{2})S_{\star F\bar{\psi}}(u, \check{v})}\, ,\\
&P_{\psi\psi}(u|\check{v})^2 =- \frac{{\sqrt{g x y}}\,S_{\psi\psi}(u, \check{v})}{(xy-g^2)(u-v+i)S_{\star \psi\psi}(u, \check{v})}\, ,\\
&P_{\psi\bar{\psi}}(u|\check{v})^2 = -\frac{\,(xy-g^2)S_{\psi\bar{\psi}}(u, \check{v})}{{\sqrt{g x y}}\,(u-v) S_{\star \psi\bar{\psi}}(u, \check{v})}\, ,\\
&P_{\psi\psi}(\check{u}|\check{v})^2 = \frac{(xy-g^2)S_{\psi\psi}(\check{u}, \check{v})}{{\sqrt{g xy}\,}(u-v)(u-v+i)S_{\star \psi\psi}(\check{u}, \check{v})}\, ,\\
&P_{\psi\bar{\psi}}(\check{u}|\check{v})^2 = \frac{{\sqrt{g xy}\,}S_{\psi \bar{\psi}}(\check{u}, \check{v})}{(xy-g^2)S_{\star \psi \bar{\psi}}(\check{u}, \check{v})}\, ,\\
\end{aligned}
\eeq
where, again, `check marked' rapidities indicate analytical continuation to the small momentum sheet.

More explicitly, and including bound states as well, we can take all the transitions listed before and perform the continuation. The explicit form of the analytically continued transitions still preserves the structure (\ref{FandExponent}), but the prefactors as well as the functions in the exponent are changed. The continuation of the exponent was explicitly worked out in \cite{2pt} and here we provide once again only the expressions for the prefactors. We stress in particular that $f_{\psi_{S}}(u) \neq f_{\psi}(\check{u})$ and $F_{X\psi_S}(u,v)\neq F_{X\psi}(u,\check{v})$, since upon continuation of the \textit{full} transition some extra terms are produced by the exponent in~(\ref{FandExponent}) and transferred to the prefactors. Instead, the correct analytic continuation produces $f_{\psi_{S}}(u) = 1$ and
\begin{equation}
\begin{aligned}
&F_{F_a\psi_S}(u,v) = F_{\psi_S F_a}(-v, -u) = -\frac{\sqrt{g}(u-v+i a/2)}{y(x^{[+a]}x^{[-a]})^{1/4}\sqrt{(1-\frac{g^2}{x^{[+a]}y})(1-\frac{g^2}{x^{[-a]}y})}} \,,\,\,\,\,\,\,\, \text{for }a>0\,,\\
&F_{F_{a}\psi_S}(u,v) = F_{\psi_S F_{a}}(-v,-u) = \frac{(x^{[+a]}x^{[-a]})^{1/4}}{\sqrt{g}}\sqrt{\big(1-\frac{g^2}{x^{[+a]}y}\big)\big(1-\frac{g^2}{x^{[-a]}y}\big)}\,,\,\,\,\,\,\,\, \text{for }a<0\,,\end{aligned}
\end{equation}
and
\begin{equation}
\begin{aligned}
&F_{\phi\psi_S}(u,v) =  1/\sqrt{y}, \qquad \qquad \qquad \,\,\,\,\,\,\,\,\, F_{\psi_{S}\phi}(u, v) = 1/\sqrt{x}\, ,\\
&F_{\psi\psi_S}(u, v) = \frac{g^{1/4}}{{ x^{1/4}y^{3/4}}}/\sqrt{1-\frac{g^2}{xy}}\, , \qquad\,\,  F_{\psi\bar\psi_S}(u, v) =  -\frac{1}{g^{1/4}} \left(\frac{x}{y}\right)^{1/4} \sqrt{1-\frac{g^2}{xy}}\, ,  \\
&F_{\psi_{S}\psi}(u, v) = -\frac{g^{1/4}}{{x^{3/4}y^{1/4}}}/\sqrt{1-\frac{g^2}{xy}}\, , \qquad\,\,  F_{\psi_{S}\bar\psi}(u, v) =  -\frac{1}{g^{1/4}} \left(\frac{y}{x}\right)^{1/4} \sqrt{1-\frac{g^2}{xy}}\, ,  \\
&F_{\psi_S \psi_S}(u,v) = -\frac{(xy)^{1/4}}{g^{1/4}(u-v)} \sqrt{1-\frac{g^2}{xy}}\, ,   \qquad F_{\psi_S \bar\psi_S}(u, v)= -\frac{g^{1/4}}{(xy)^{1/4}}/\sqrt{1-\frac{g^2}{xy}}\,.
\end{aligned}
\end{equation}

\subsection{Measures}\la{measures}
We recall that the measures are obtained from the direct transitions through
\beq
\text{Res}_{v=u}\,P_{X|X}(u|v)=\frac{i}{\mu_{X}(u)}\,.
\eeq
They have the universal structure
\beq
\mu_{X}(u) = \frac{M_{X}(u)}{f_{X}(u)f_{X}(-u)}  \exp\Big[2\tilde\kappa_X(u)^t\cdot \mathcal{M} \cdot\tilde\kappa_X(u)-2\kappa_X(u)^{t}\cdot \mathcal{M} \cdot\kappa_X(u)\Big]\, ,
\eeq
with
\begin{equation}
\begin{aligned}
M_{\phi}(u)&=\frac{\pi  {g}}{\cosh(\pi  u)}\,, \\
M_{\psi}(u)&=-i\frac{\pi g^{5/4} u}{{\sqrt{x}} \sinh (\pi  u) \sqrt{x^2-g^2}} \, , \\
M_{F_a}(u)&=\frac{(-1)^ag^2\Gamma(1+\frac{a}{2}+iu)\Gamma(1+\frac{a}{2}-iu)}{\Gamma(a)(x^{[+a]}x^{[-a]}-g^2)\sqrt{((x^{[+a]})^2-g^2)((x^{[-a]})^2-g^2)}}\,,\,\,\,  \textrm{for}\,\, a >0\,.
\end{aligned}
\end{equation} 
Moreover, we have that $\mu_{F_{-a}}(u) = \mu_{F_a}(u)$ and $\mu_{\bar{\psi}}(u) = -\mu_{\psi}(u)$. Upon analytical continuation to the small fermion sheet, we obtain
\beq
M_{\psi_S}(u)=i\frac{{g^{1/4}}{\sqrt{x}}}{\sqrt{x^2-g^2}}\,.
\eeq

\subsection{Zero momentum limit}\label{zmf}

Given a transition $P_{\bar{\psi}|Y}(\check{u}|v)$ it is immediate to derive its scalings at $u = \infty$, i.e.~for a zero momentum fermion. This one can be read directly from the function $F_{\bar{\psi}_{S}Y}(u, v)$ listed before, since the remaining factors in~(\ref{FandExponent}) all go to $1$ in this limit. We get this way
\beq\label{zerof}
\begin{aligned}
& P_{\bar{\psi}|\bar{\psi}}(\check{u}|v) \sim \frac{g^{1/4}}{{u^{3/4}y^{1/4}}}\, , \\
& P_{\bar{\psi}|\phi}(\check{u}|v) \sim \frac{1}{\sqrt{u}}\, , \\
& P_{\bar{\psi}|\psi}(\check{u}|v) \sim \frac{1}{g^{1/4}} \left(\frac{y}{u}\right)^{1/4} \, , \\
& P_{\bar{\psi}|F_a}(\check{u}|v) \sim \frac{\sqrt{g}}{(y^{[+a]}y^{[-a]})^{1/4}} \,,\,\,\,\,\,\,\, \text{for }a<0\,,\\
& P_{\bar{\psi}|F_{a}}(\check{u}|v) \sim \frac{(y^{[+a]}y^{[-a]})^{1/4}}{\sqrt{g}}\,,\,\,\,\,\,\,\, \text{for }a>0\, ,
\end{aligned}
\eeq
where, again, $y = x(v)$ and $y^{[\pm a]} = x(v\pm i\frac{a}{2})$. This information was used to obtain the non-MHV form factors $h_{\bar{Y}}(v)$ of the excitation $Y(v)$ in the bulk of the paper, with help of the asymptotic behaviour of the Jacobian factor
\beq
\sqrt{\frac{\Gamma_{\textrm{cusp}}}{2ig}\mu_{\bar{\psi}}(\check{u}) \frac{du}{dp_{\bar{\psi}}}(\check{u})} \sim \frac{u^{3/4}}{g^{3/8}}\, ,
\eeq
itself following from
\beq
\mu_{\bar{\psi}}(\check{u}) \sim -\frac{ig^{1/4}}{\sqrt{u}}\, , \qquad \frac{dp_{\bar{\psi}}}{du}(\check{u}) \sim -\frac{\Gamma_{\textrm{cusp}}}{2u^2}\, .
\eeq

\section{The superconformal charge ${\cal Q}$ and the flux Goldstone fermion}\la{Qcommutator}

In section \ref{sec2point2} we used the realization of the superconformal generator ${\cal Q}_A$ as a zero momentum fermion. The precise relation is (\ref{Qtopsi}) and is repeated here for convenience
\beq\la{Qtopsi2}
{\cal Q}|0\>=\sqrt{\Gamma_\text{cusp}\over 2g}\lim_{p\to0}|p\>=\lim_{v\to\infty}\sqrt{{\Gamma_\text{cusp}\over2 gi}\,\frac{d\check{v}}{dp_{\bar\psi}}\mu_{\bar\psi}(\check v)}\,|\bar{\psi}(\check v)\>\ .
\eeq
In this appendix we will derive this relation. 

\subsection{The zero momentum fermion}\la{B1}

The non-trivial part of (\ref{Qtopsi2}) is the factor dressing the zero momentum fermion state. To derive it, we should first fix the normalization of that flux-tube state $|\bar\psi\>$. It is instructive to do this in two steps. First we note that we have a well defined flux tube square measure $\mu$, which allows us to overlap states in the flux Hilbert space. Using the supersymmetry algebra, this measure leads to a precise representation of the superconformal generators on the flux, that we denote by $\mathbb{Q}_A$. There is no reason however for this realization of the supercharge on the flux tube to be normalized in the same way as its realization on the generating function of amplitudes or equivalently, the super loop. Namely, the two may differ by an overall proportionality constant ${\mathbb Q}_A=c_0\times {\cal Q}_A$.\footnote{See \cite{superloopsimon} for a very similar relation.} We shall now first relate ${\mathbb Q}_A$ to a zero momentum fermion using the measure and the supersymmetry algebra. We will then fix the constant $c_0$ by demanding that pentagon NHMV amplitude is the same as the MHV one as appears in the generating function (\ref{superloopinPs}).

Consider first a delta-function normalized momentum states
\beq\la{pnorm}
\<p(u)|\tilde p(v)\>=2\pi\delta(p(u)-\tilde p(v))\qquad\Rightarrow\qquad \<p=0|\tilde p=0\>=2\pi\delta(0)= \text{Vol}(\sigma)
\eeq
where $\text{Vol}(\sigma)$ is the infinite volume of the flux in the coordinate $\sigma$ conjugate to $p$. These momentum states differ by a simple normalization factor (involving the measure $\mu_{\bar\psi}$) from the rapidity states, which we conventionally normalize as \cite{data}
\beq
\<v|u\>={2\pi\over\mu_{\bar\psi}(u)}\delta(v-u)\qquad\Rightarrow\qquad|p(u)\>=\sqrt{-i\frac{d v}{dp_{\bar\psi}}\mu_{\bar\psi}(u)}\,|u\>
\eeq
On the other hand, with respect to this square measure, we have
\beq\la{Qnormpre}
2||{\mathbb Q}_A|0\>||^2=
\<0|\{{\mathbb Q}_A,\bar{\mathbb Q}^A\}|0\>
\eeq
where $|0\>$ is the GKP vacuum, normalized so that $\<0|0\>=1$. The commutator is a special conformal generator that can be written in terms of the symmetries of the square as 
\beq\la{commutation}
\sum_{A=1}^4\{{\mathbb Q}_A,\bar{\mathbb Q}^A\}=2(\d_\tau-i\d_\phi)+{\cal C}\ ,
\eeq
where the total helicity ${\cal C}=0$ in our case and for simplicity, we have summed over the R-charged index, see section \ref{B2}. The GKP vacuum does not carry $U(1)$ charge while $\d_\tau$ measures its energy. We conclude that
\beq\la{Qnorm}
||{\mathbb Q}_A|0\>||^2={1\over4}E_\text{GKP}\<0|0\>={1\over4}\Gamma_\text{cusp}\,\text{Vol}(\sigma)
\eeq
where in the last step we used the interpretation of $\Gamma_\text{cusp}$ as the energy density of the flux in the $\sigma$ direction \cite{AldayMaldacena}. It then follows that
\beq\la{Qtopsi3}
{\mathbb Q}|0\>=\sqrt{\Gamma_\text{cusp}\over 4}\lim_{p\to0}|p\>=\lim_{v\to\infty}\sqrt{{\Gamma_\text{cusp}\over4 i}\,\frac{d\check{v}}{dp_{\bar\psi}}\mu_{\bar\psi}(\check v)}\,|\bar{\psi}(\check v)\>
\eeq
where the check over the fermion rapidity ($\check v$) indicates that it is on the so-called {\it small fermion sheet} where the zero momentum point is, see \cite{2pt}. 

Next, we shall fix the proportionality constant $c_0$. This is done by demanding that the pentagon NHMV amplitude is the same as the MHV one. Translated to the POPE notations, this condition reads
\beqa
1\!\!\!&=&\!\!\!P(0|0)
{\stackrel{!}{=}}P^{[4]}(0|0)\\
\!\!\!&=&\!\!\!{1\over c_0^4}\({\Gamma_\text{cusp}\over4}\)^2\prod\limits_{j=1}^4\,\lim_{v_j\to\infty}\sqrt{-i\frac{d\check v_j}{dp_{\bar\psi}}\mu_{\bar\psi}(\check v_j)}\times P_{\bar\psi^4|0}(\check v_1,\check v_2,\check v_3,\check v_4|0)\times (\texttt{matrix part})= {g^2\over 4 c_0^4}\nn
\eeqa
where in the last step we used the large $v$ behaviours quoted in~\ref{zmf}, together with
$$P_{\bar \psi|\bar\psi}(\check v|\check u)\sim{(vu)^{1\over4}\over g^{1\over4}(u-v)}\, ,$$
that can be read from the expressions in~\ref{APPsmall}, and, finally, the expression for the matrix part (which is nontrivial in this case) given by~\cite{FrankToAppear}
$$\texttt{matrix part}=\frac{1}{\prod\limits_{i>j}^4(v_i- v_j+i)}\, .$$
We deduce that $c_0=\sqrt{g/2}$ and hence the relation (\ref{Qtopsi2}). In the following subsection we elaborate on the commutation relation (\ref{commutation}). 

\subsection{The commutator of superconformal charges}\la{B2}

We shall now derive the relation relation (\ref{commutation}) used above. 
For that aim, it is convenient to decompose any twistor in the basis of the square four twistors, (see for example appendix A of \cite{data} for an explicate choice)
\beq\la{squarebase}
Z=z_b\,Z_\text{bottom}+z_t\,Z_\text{top}+z_r\,Z_\text{right}+z_l\,Z_\text{left}\ .
\eeq
In this basis, the three symmetries of the square are generated by
\beq
\d_\tau=z_b\d_{z_b}-z_t\d_{z_t}\ ,\qquad\d_\sigma=z_r\d_{z_r}-z_l\d_{z_l}\quad\text{and}\quad\d_\phi={i\over2}\(z_b\d_{z_b}+z_t\d_{z_t}-z_r\d_{z_r}-z_l\d_{z_l}\)\,.
\eeq
The relation (\ref{commutation}) is an algebra relation between superconformal generators and therefore we can use any representation of the generators to test it. When acting on the generating function of helicity amplitudes, the supercharge is represented as $Q_A^{\alpha}=Z^\alpha\d_\eta$. The operator ${\cal Q}_A$  is a specific component of the superconformal generator $Q_A^{\alpha}$ 
which was specified in \cite{shortSuper}. To translate between (\ref{squarebase}) and the notations of \cite{shortSuper} we may think of the square here as the bottom square of the $j$'th pentagon in the POPE decomposition. Then, equations (10)-(11) in \cite{shortSuper} for the component of $Q_A^{\alpha}$ read
\beq\la{Qcomp}
{\cal Q}_A=\d_{\chi^A_j}\ \propto\  \(Z_{j-1}\wedge Z_j\wedge Z_{j+1}\)\cdot Z\d_\eta\ \propto\ z_b\d_{\eta^A}
\eeq
where we used that $Z_j=Z_\text{left}$, $Z_{j-1}=Z_\text{right}$ and $Z_{j+1}$ is a linear combination of $Z_\text{right}$ and $Z_\text{top}$. Here, we drop the proportionality factors in (\ref{Qcomp}) as it drops out in the commutator (\ref{commutation}). We can now use the conjugate operator in this representation to evaluate the commutator in (\ref{commutation}). We find\footnote{Note that the commutation relation (\ref{commrelation}) is independent of the measure one uses to realise it and thus $\{{\cal Q}_A,\bar{\cal Q}^A\}=\{{\mathbb Q}_A,\bar{\mathbb Q}^A\}$.}
\beqa\la{commrelation}
\sum_A\{{\cal Q}_A,\bar{\cal Q}^A\}&=&\sum_A\{z_b\d_{\eta^A},\eta^A\d_{z_b}\}=4z_b\d_{z_b}+\sum_A\eta^A\d_{\eta^A}\\&=&2(\d_\tau-i\d_\phi)+\(z_b\d_{z_b}+z_t\d_{z_t}+z_r\d_{z_r}+z_l\d_{z_l}\)+\sum_A\eta^A\d_{\eta^A}\nn\\
&=&2(\d_\tau-i\d_\phi)+{\cal C}\nn
\eeqa
Here, the summation over the R-charge index was done for simplicity. Otherwise, on the right hand side we would also had an R-charge generator. Alternatively to this derivation, (\ref{commutation}) can be read from equation (3.9) in \cite{Korchemsky:2010ut} by specifying to the corresponding component.

\section {A second example: $\cP_1\circ\cP_2\circ\cP_{34}$ } \la{anotherEx}

In this second example we analyse another heptagon component where the middle pentagon is charged, namely $\cP_1\circ\cP_2\circ\cP_{34}$. We consider again the contribution from multiparticle states in both squares, this time a term proportional to
\beq
e^{-2\tau_1-3i\phi_1/2}\times e^{-2\tau_2+i \phi_2}\la{ex2prop}\,.
\eeq
To find which processes contribute we first note that we need twist 2 states for both squares with helicities -3/2 and +1. Then we follow the same logic as in section~\ref{NMHVhep} and see the implications of $R$-charge conservation.
The first pentagon carries one unit of $R-$charge which enforces the state propagating in the first middle square to be in the anti-fundamental ($\bar{\bf{4}}$) representation of $SU(4)$. The second pentagon also possesses one unit of $R-$charge. This implies that the state in the second middle square should be in one of the two representations entering in the decomposition of $\bar{\bf{4}}\otimes\bar{\bf{4}}=\bf{6}\oplus \overline{\bf{10}}$. The state is finally projected onto the vector representation ($\bf{6}$) by the last pentagon that carries two units of $R-$charge. In the first square, the only option is the state formed by $\bar\psi F_{-1}$. In the second one we could either have $\phi F_1$ or $\psi\psi$; as the reader might guess, it is the second state which contributes at tree level given the small fermions behaviour. All we are left to do is study the process
\beq
\texttt{vacuum}\rightarrow\bar\psi \,F_{-1}\rightarrow\psi \psi\rightarrow \texttt{vacuum}\,.
\eeq
We see that the R-charge index of the antifermion in the first middle square is fixed by the R-charge index of $\chi$ in the first pentagon. However, this is not the case for the two fermions in the second square. In other words the matrix part is not trivial in this case, but one can compute it in a straightforward way following the logic in \cite{2pt} where the contribution of $(\psi\bar\psi)$ was studied and the S-matrix of the fermions in the vector ({\bf 6}) representation  \cite{Belitsky:2014sla}. The full integrand is then given by the product of the following parts
\beqa
\texttt{dynamical part}&=&\hat\mu_{F_{-1}}(u)\,\hat\mu_{\bar\psi}(w_1)\,\hat\mu_{\psi}(w_2)\,\hat\mu_{\psi}(v)\\
&&\times \frac{P_{\bar\psi|F_{1}}(w_2|u)P_{\bar\psi|F_{1}}(v|u)P_{\bar\psi|\psi}(w_2|w_1)P_{\bar\psi|\psi}(v|w_1)}{P_{F_{1}|\psi}(u|w_1)P_{\bar\psi|F_{-1}}(w_1|u) P_{\psi|\psi}(w_2|v)P_{\bar\psi|\bar\psi}(v|w_2)}\,,\\
\texttt{form factors part}&=&\frac{1}{{g^{\frac{5}{4}}}} h_{\psi}(w_1)h_{\bar\psi}(w_1) h_{F_{-1}}(u) h_{\bar\psi}(w_2) h_{\bar\psi}(v)(h_{\psi}(w_2))^2 (h_{\psi}(v))^2,  \\
\texttt{matrix part}&=& \frac{2}{(v-w_2)^2+1} \,.
\eeqa
Given that the fermions can be small or large, one can split the process into the different contributions
\beqa\label{terms}
\mathcal{W}_{\bar\psi \,F_{-1}\rightarrow\psi \psi} &=&\,\mathcal{W}_{\bar\psi_L \,F_{-1}\rightarrow\psi_L \psi_L}+\mathcal{W}_{\bar\psi_L \,F_{-1}\rightarrow\psi_S \psi_S}+2\times \mathcal{W}_{\bar\psi_L \,F_{-1}\rightarrow\psi_S \psi_L}\\
&&+\,\mathcal{W}_{\bar\psi_S \,F_{-1}\rightarrow\psi_L \psi_L}+\mathcal{W}_{\bar\psi_S \,F_{-1}\rightarrow\psi_S \psi_S}+2\times \mathcal{W}_{\bar\psi_S \,F_{-1}\rightarrow\psi_S \psi_L}\,,
\eeqa
where the factor of $2$ accounts for the two equivalent ways of choosing which of the two fermions is small. At tree level only the last term contributes. One could expect that also the second to last term with two small fermions in the last square would also appear at tree level, but as explained in \cite{2pt} this contribution vanishes\footnote{This is because after integrating out the first small fermion there are no more singularities enclosed by the second integration contour. In colloquial words, one should have at least one large excitation to attach the fist small fermion (e.g.~$F_1\psi_S\psi_S$ would have a non vanishing contribution).}. Therefore we consider only the last term in (\ref{terms}) and the transition reads
\beqa
\mathcal{W}_{\bar\psi \,F_{-1}\rightarrow\psi \psi}&=&2\int\limits_{\mathbb{R}} \frac{du}{2\pi} \int\limits_{{\mathbb{R}-i0}} \frac{dv}{2\pi} \int\limits_{{\bar{\mathcal{C}}_\text{small}}}\frac{dw_2}{2\pi}\int\limits_{\mathcal{C}_\text{small}} \frac{dw_1}{2\pi}\,  (\texttt{dynamical part})\nn\\
&&\times\, (\texttt{form factors part})\times(\texttt{matrix part})\,,
\eeqa
where {we use $-i\epsilon$ prescription for fermions in the top square and the opposite prescription for fermions in the bottom square. (Accordingly we use the lower half plane (half-moon) contour $\mathcal{C}_\text{small}$  for the small fermion at the bottom and its conjugate $\bar{\mathcal{C}}_\text{small}$ for the small fermion at the top.)} We can now integrate out the two small fermions by picking the poles ${w_2=v+i}$ coming from the pentagon transition $P_{\bar{\psi}F_{-1}}$ and $w_1=u-i/2$ from the matrix part.
Finally we integrate over $u$ and $v$ and find that the tree-level prediction is given by
\beqa
\mathcal{W}_{\bar\psi \,F_{-1}\rightarrow\psi \psi}&=&\frac{e^{-2\tau_1-2\tau_2-3i\phi_1/2+i\phi_2}}{\left(e^{2 \sigma _1}+1\right){}^2 \left(e^{2 \sigma _2}+1\right){}^2 \left(e^{2 \sigma _1}+e^{2 \sigma _2}+e^{2 \sigma _1+2 \sigma _2}\right){}^3}\nn\\
&&\times\(e^{7 \sigma _1}+3 e^{5 \sigma _1+2 \sigma _2}+5 e^{7 \sigma _1+2 \sigma _2}+3 e^{3 \sigma _1+4 \sigma _2}+12 e^{5 \sigma _1+4 \sigma _2}\right.\nn\\
&&\left.\,\,\,\,\, +10 e^{7 \sigma _1+4 \sigma _2}+e^{\sigma _1+6 \sigma _2}+5 e^{3 \sigma _1+6 \sigma _2}+10 e^{5 \sigma _1+6 \sigma _2}+6 e^{7 \sigma _1+6 \sigma _2}\) \label{secresult}\,.
\eeqa
To check this expression against data, we first relate $\cP_1\circ\cP_2\circ\cP_{34}$ to a linear combination of super amplitude $\eta$-components \cite{shortSuper}
\beqa
\frac{\partial}{\partial\chi_1}\frac{\partial}{\partial\chi_2}\(\frac{\partial}{\partial\chi_3}\)^2\mathcal{W}\!\!&=&\!\!\dfrac{(-\textbf{1})_1 \((\textbf{5})_3\)^2}{(\textbf{1})_2(\textbf{2})_2(\textbf{3})_2 } \dfrac{\partial}{\partial\eta_{-1}} \(\<1,2,3,-1\>\dfrac{\partial}{\partial\eta_{-1}}+\<1,2,3,0\>\dfrac{\partial}{\partial\eta_{0}}\) \(\dfrac{\partial}{\partial\eta_5}\)^2\mathcal{W} \nonumber\\
\!\!&=&\!\!\dfrac{(-\textbf{1})_1 \((\textbf{5})_3\)^2}{(\textbf{1})_2(\textbf{2})_2(\textbf{3})_2} \(\<1,2,3,-1\>\,\mathcal{W}^{(-1,-1,5,5)}+\<1,2,3,0\>\,\mathcal{W}^{(-1,0,5,5)}\).
\eeqa 
Then we extract the amplitude components in Mathematica and evaluate the full expression with the heptagon twistors. After expanding at large $\tau_1$ and $\tau_2$ and picking up the term proportional to \eqref{ex2prop} we find again a perfect match with the OPE result \eq{secresult}.

\end{document}